\documentclass[preprint]{aastex6}

\usepackage{graphicx}
\usepackage[space]{grffile}
\usepackage{latexsym}
\usepackage{amsfonts,amsmath,amssymb}
\usepackage{url}
\usepackage{hyperref}
\hypersetup{colorlinks=false,pdfborder={0 0 0}}
\usepackage{textcomp}
\usepackage{longtable}
\usepackage{multirow,booktabs}
\usepackage[english]{babel}
\usepackage{rotating}

\usepackage{epstopdf}
\shorttitle{Investigation of the Interaction Between MIR Bubbles and the ISM}
\shortauthors{Kathryn Devine}

\begin{document}

\title{A Molecular Line Investigation of the Interaction Between Mid-Infrared Bubbles and the Interstellar Medium}

\author{Kathryn Devine}
\affil{The College of Idaho}
\email{KDevine@collegeofidaho.edu}

\author{Johanna Mori}
\affil{The College of Idaho}
\email{johannamori1@gmail.com}

\author{Christer Watson}
\affil{Manchester University (formerly Manchester College)}
\email{cwatson@manchester.edu}

\author{Leonardo Trujillo}
\affil{The College of Idaho}
\email{leonardo.trujillo.m@gmail.com}

\author{Matthew Hicks}
\affil{Manchester University (formerly Manchester College)}
\email{MTHicks2017@spartans.manchester.edu}

\received{2017 June 27}
\revised{2018 April 6}
\accepted{2018 May 14}
\published{2018 July 11}

\begin{abstract}
We used the Green Bank Telescope to detect molecular lines observed toward Mid-Infrared (MIR) bubbles N62, N65, N90, and N117. The bubbles were selected from \citet{Watson2016} who detected non-Gaussian CS (1-0) emission lines toward the bubbles. Two of the bubbles are adjacent to infrared dark clouds (IRDCs); we examined these sources for evidence of interaction between the bubble rim and IRDC. The other two bubbles contain YSOs interior to the bubble rim; in these sources we observed the gas near the YSOs. We detect CS (1-0) emission toward all of the sources, and in several pointings the CS emission shows non-Gaussian line shapes. HC$_3$N (5-4), C$^{34}$S (1-0), CH$_3$OH (1-0), and SiO (v=0) (1-0) were also detected in some pointings. We calculate column densities and abundances for the detected molecules. We compare the velocity of optically-thick CS emission with the velocity of the other, optically thin lines to look for evidence of infall.  We find that even in pointings with non-Gaussian CS emission, our detections do not support an infall model. We interpret the kinematics of the gas in N62, N65, and N117 as likely evidence of multiple clouds along the line of sight moving at slightly offset velocities.  We do not detect evidence of bubble rims interacting with IRDCs in N62 or N90. The gas interior to bubbles appears more disrupted than the gas in the IRDCs. N65 shows significantly stronger emission lines than the other sources, as well as the most complicated non-Gaussian line shapes.
\end{abstract}

\section{Introduction}
Young, massive stars are thought to trigger star formation by injecting energy into the interstellar medium (ISM). However, massive stars are found in areas of active star formation and thus it is difficult to determine whether a massive star caused the formation of nearby stars via triggering or whether the stars are simply all forming together as part of the same star forming region. Mid-infrared (MIR) bubbles represent a class of objects that could contain observable evidence of triggering by massive stars.  Discovered in MIR Galactic surveys, MIR bubbles are formed by young, massive stars, and probing their interaction with the surrounding ISM might resolve whether bubbles' expansion induces collapse in their surroundings. In this paper, we examine four MIR bubbles for evidence of their interaction with the ISM to probe whether these interactions support a triggering scenario.

MIR bubbles, named for their morphology in mid-infrared images, have been extensively studied with the aim of understanding both their structure and influence on the surrounding interstellar medium \citep{Churchwell2006,Watson2008,Deharveng2010,Simpson2012,Kendrew2012,Kendrew2016}.  A defining characteristic of MIR bubbles is an 8 $\mu$m rim of glowing poly-cyclic aromatic hydrocarbon (PAH) molecules that have been excited by ionizing ultraviolet photons.  This rim traces the boundary between a warm, ionized interior and exterior molecular gas \citep{Churchwell2006,Churchwell2007}.  While other objects may take on a bubble-like morphology in the MIR (e.g. supernova remnants or planetary nebulae), a majority of bubbles contain 24 $\mu$m emission tracing warm dust and $20$ cm radio continuum emission indicating the presence of ionized gas \citep{Watson2008,Watson2009}. These characteristics are consistent with most MIR bubbles being expanding H {\small II} regions around young OB stars \citep{Watson2009, Deharveng2010}.

The extent and mechanisms connecting bubble expansion with subsequent star formation are not yet understood. Triggered star formation mechanisms, such as collect-and-collapse \citep{Elmegreen1977} and radiatively-driven implosion \citep{Bertoldi1989} have been used to explain observations of gas and YSOs surrounding bubbles, but with some controversy \citep{Samal2014,Watson2010,Ortega2013}. One piece of evidence often cited in support of triggering is an overabundance of YSOs coincident with bubble rims.  For example, \citet{Watson2010} used 2MASS \citep{Skrutskie2006} and GLIMPSE \citep{Benjamin2003} photometry and spectral energy diagram (SED) fitting to identify YSO candidates toward $46$ MIR bubbles, and found roughly a quarter of the bubbles in the sample showed such a YSO overabundance.  \citet{Watson2016} conducted follow-up CS (1-0) observations of the dense gas toward $44$ YSOs in $10$ of the \citet{Watson2010} bubbles to further explore the association between the bubbles and YSOs. \citet{Watson2016} detected CS in $18$ of the $44$ pointings, with four of the bubbles containing CS with non-Gaussian line shapes.  Three of these non-Gaussian profiles had signs of self-absorption with increased blue-shifted emission possibly indicative of infall and thus star-formation, while the fourth was explained by multiple cloud components along the line of sight.  The interpretation of the \citet{Watson2016} CS data, however, was limited by the lack of observations of additional, optically-thin dense gas tracers.  

In this study, we present follow-up observations toward the four \citet{Watson2016} bubbles with non-Gaussian CS (1-0) line profiles.  These observations probe the nature of the observed line profiles, which in turn allows us to examine whether gas  exterior and interior to the bubble rims show signs of infall associated the expanding bubble. We observed numerous molecular line transitions between $38-50$ GHz using the National Radio Astronomy Observatory's (NRAO) \footnote{The National Radio Astronomy Observatory is a facility of the National Science Foundation operated under cooperative agreement by Associated Universities, Inc.} Green Bank Telescope (GBT). Two of the bubbles, N62 and N90, have infrared dark clouds (IRDCs) exterior to but coincident with their rims. Since IRDCs are believed to contain the conditions necessary for star formation, we sought to relate the possible infall detected in \citet{Watson2016} to interactions between the bubble rims and the IRDCs. Evidence of infall only where the bubble rim and IRDC are coincident would support a triggering scenario. The other two bubbles, N65 and N117, contain molecular gas clumps and YSOs interior to the bubble rim. In these sources, we examined the kinematic properties (velocities, linewidths, and evidence of infall) of the gas to see how the gas may have been influenced by the expanding bubble rim.

To relate features traced in the MIR with observations of molecular gas we must verify a physical association between the molecular gas and IR features, rather than line-of-sight coincidence.  The l radio observations of the bubbles presented in this study by \citet{Watson2016} used a wide spectral window of $\sim 150$ km s$^{-1}$.  CS lines at only one velocity were detected toward the sources, indicating that there are not several molecular sources along the line of sight to the IR features.  We discuss the correlation of the velocity of the gas we observed with previously reported velocities for the bubbles in Section \ref{sec:sources}, in which we verify that our dense-gas observations of N62, N65, and N117 are likely associated with the MIR features. In N65 and N117, we have assumed that the gas and YSOs are interior to the bubbles, although we consider alternative geometries in Section \ref{sec:discussion}.

We elaborate on the sources selected in this study in Section \ref{sec:sources}, describe our observations in Section \ref{sec:observations}, and report detections and line fitting results in Section \ref{sec:results}.  We calculate column densities and combine these column densities with \emph{Herschel} Hi-GAL data to determine abundances in Section \ref{sec:abundances}.  We examine the line kinematics in Section \ref{sec:kinematics}.  We discuss kinematic trends and our interpretation of these trends in Section \ref{sec:discussion}.  Our conclusions are summarized in Section \ref{sec:conclusions}.

\section{Sources}
\label{sec:sources}

The four bubbles in this study were selected from the \citet{Watson2016} sample based on non-Gaussian line profiles seen in CS (1-0) emission, indicating interesting kinematics in the gas near the bubble rims.  The bubbles are shown in Figures \ref{fig:bubbles} and \ref{fig:N65fig}.  Two of the bubbles, N62 and N90, had rims coincident with IRDCs.  In these cases, observations were taken along the length of the IRDC.  The other two bubbles, N65 and N117, contain YSOs and dense gas interior to the bubble rim. In these bubbles, the observations sampled the YSO and gas around the YSO.  The Galactic longitude and latitude of the pointings are given in Table \ref{tab:pointings}, and the pointings are indicated by the circles in  Figs. \ref{fig:bubbles} and \ref{fig:N65fig}. 

\subsection{N62}

N62 was included in the sample of MIR bubbles studied by \citet{Deharveng2010}, who noted that it was coincident with 21 cm radio emission, detected in the ATLASGAL survey, and coincident with an infrared dark cloud (IRDC). N62 is coincident with H{\small II}  region G34.325+0.211 \citep{anderson2011}, and \citet{xu2016} conducted an analysis of the ionizing flux from this H{\small II} region to determine that the spectral type of the ionizing star is between B1V and B1.5V. The  H{\small II} region velocities reported by  \citep{anderson2011} are consistent with the CS (1-0) velocities of the dense gas detected in \citet{Watson2016} and in this paper, indicating that the IRDC is indeed at the same distance as, and associated with, N62.

\citet{xu2016} examined the association between N62 and IRDC G34.43+0.24.  \citet{xu2016} conducted a multiwavelength study toward the IRDC and suggest it contains massive protostars, UC H{\small II} regions, and six infrared bubbles, including N62. Using CO emission, \citep{xu2016} determined the IRDC filament has a velocity 53-63 km s$^{-1}$, consistent with the velocity range of the gas detected in this paper.  Interestingly, \citet{xu2016} detected $^{13}CO J=1-0$ and $C^{18}O J=1-0$ lines toward H{\small II} complex G34.26+0.15 (~5 arcmin south of N62) with double peaked, asymmetric profiles like those seen in N62 by \citet{Watson2016}; like \citet{Watson2016}, \citet{xu2016} concluded these profiles may be the result of infall or accretion. \citet{xu2016} suggest that IRDC G34.43+0.24 is an extended ($\sim 37$ pc) IRDC that contains three regions, each hosting different stages of star formation.  N62 is located in the region that they classify as hosting early star formation. 

\subsection{N65}
N65 and indicators of star formation coincident with the bubble have been well studied in previous research.  N65 was reported in both the \citet{Churchwell2006} and \citet{Simpson2012} surveys of bubbles.  An H{\small II} region coincident with N65 is reported by \citet{Andersonbania2009} at the same velocity as the CS (1-0) detections reported in \citet{Watson2016} and in this study, indicating that the dense gas we detect is associated with the bubble. \citet{Deharveng2010} included N65 in their mid-IR and radio continuum bubble analysis, and in their sub-sample of 16 objects with ultra-compact H{\small II} regions for which they conducted a more detailed analysis.  \citet{Deharveng2010} report an ATLASGAL 870 um continuum peak coincident with the pointings used in this study associated.  The mass of this condensation is estimated to range between 560-2000 M$_{solar}$ (\citet{Deharveng2010} and references therein).  The 870 um peak is also near an UC H{\small II} region \citep{kurtz1994, kurtz2005} and several methanol masers (\citet{cyganowski2009}, see discussion below); the location of these sources relative to the mid-IR bubble is shown in Fig.  \ref{fig:N65fig}. \citet{Deharveng2010} interpreted the 870 um dust condensation as a sign of compression between the rim of N65 and an adjacent bubble, N65bis, and suggest that the signs of star formation in the region presented in this study are consistent with young massive stars forming at the interface between two expanding bubbles.  

\citet{cyganowski2009} report an “extended green object (EGO)” with bipolar morphology coincident with our observations. Their survey of methanol masers toward the EGO show a 6.7 GHz Class II maser at the center of the EGO, coincident with a 24 um peak, and a chain of Class I 44 GHz masers offset from the EGO (Fig.  \ref{fig:N65fig}).  In follow-up work, \citet{cyganowski2011} observed the EGO at radio cm continuum emission with $\sim$1 arcsecond resolution and found 5 compact continuum sources (shown in Fig.  \ref{fig:N65fig}).  \citet{forstercaswell} and \citet{argon2000} report OH and H$_2$O masers coincident with compact source CM2.  \citet{cyganowski2011} summarize that compact source CM1 likely represents a single ionizing star of spectral type B1.5V, while CM2 may represent a hypercompact H{\small II} region, and the nature of the other three cm sources remains unknown.   \citet{brogan2011} observed the \citet{cyganowski2011} compact sources at 1.3 cm with 0.05 pc resolution; they observed emission lines from NH$_{3}$ and CH$_{3}$OH, H recombination lines, and continuum.  \citet{brogan2011} detected at least four massive young stellar objects coincident with compact objects CM1 and CM2, and conclude that the region likely represents a trapezium-like protocluster.  \citet{brogan2011} also report the detection of two 25 GHz CH$_{3}$OH masers, possibly tracing shocks, which are indicated in Fig.  \ref{fig:N65fig}.

\subsection{N90}
\label{sec:N90source}

N90 was previously cataloged by \citet{Deharveng2010}, who reported several ATLASGAL condensations around the bubble rim.  They assumed a distance of 6.1 kpc, which would result in a bubble diameter of $\sim$ 14 pc.  This distance was obtained using velocities from \citet{anderson2011}, who detected an H{\small II} region at the center of N90 and detected hydrogen radio recombination lines at a $V_{LSR}$ of 70.1 km s$^{-1}$.  We note that this velocity is significantly different than the velocity at which \citet{Watson2016} detected CS emission toward the rim of N90, $\sim$35.5 km s$^{-1}$.  We hoped additional observations would better probe any potential relationship between the \citet{Watson2016} dense gas detections and the H{\small II} region; detections at both 35 and 70 km s$^{-1}$ would have fit within the bandwidth of the observations in this study (see Section \ref{sec:observations}).  We did not detect any dense gas at the same velocity as the \citet{anderson2011} radio recombination line observations.  This difference in velocity suggests that the dense gas observed is not actually associated with the H{\small II} region and bubble.  Thus, we report our emission line detections toward this source but do not provide any further analysis of these detections.

\subsection{N117}

N117 is coincident with an H{\small II} region \citep{lockman1989} and was cataloged in the \citet{Simpson2012} and \citet{Churchwell2006} bubble studies. \citet{lockman1989} reports radio recombination line velocities toward N117 of 42 km s$^{-1}$, consistent with the velocity of the CS (1-0) detections reported in \citet{Watson2016} and in this study, again indicating that the dense gas we detect is associated with the bubble.  N117 is coincident with a dense core/molecular clump \citep{rosolowsky2010} and is a source in the ATLASGAL source catalog \citep{Urquhart2014}.  Using mid-IR photometry, \citet{cyganowski2008} identified two massive stellar outflow candidates coincident with the N117 V1 and V5 pointings from this study.

\subsection{Source Distances}

We examined distances to our sources using both the \citet{reid2016} Bayesian distance calculator and previously published distances toward these sources.  The \citet{reid2016} distance calculator uses a model of the Milky Way's spiral structure mapped using parallax measurements.  This structure model is combined using a Bayesian approach with kinematic models and a source's Galactic latitude and proximity to sources with known parallax. The \citet{reid2016} calculator requires only a source's $(l, b, v)$ coordinates to assign it to a spiral arm, with good agreement between the Bayesian approach and using HI absorption to resolve the near/far distance ambiguity.  If \emph{a priori} information about the near/far distance is known, the \citet{reid2016} calculator allows users to adjust the probability of the source being located at the near distance. Using a 50\% likelihood of near/far distance and typical source velocities of N62 ($57$ km s$^{-1}$), N65 ($52$ km s$^{-1}$), N90 ($35$ km s$^{-1}$), and N117 ($40$  km s$^{-1}$), the distance calculator determined source distances of $10.55 \pm 0.29$ kpc, $10.53 \pm 0.30$ kpc, $2.68 \pm 0.39$ kpc, and $3.93 \pm 0.36$ kpc respectively.

In both N62 and N65, the \citet{reid2016} Bayesian distance calculator strongly prefers the far distance for the source based on its updated models of spiral arm structure.  The calculator places both objects at the far distance unless the calculator is given an \emph{a priori} $> 95\%$ likelihood that the bubbles are at the near distance.  Previous studies of the IRDC coincident with N62 have always assumed the near distance of $\sim 3.7$ kpc \citep{rathborne2005, xu2016}, as is common for IRDC studies.  On the other hand, \citet{Andersonbania2009} place an UC H{\small II} region coincident with N62 at the far distance (10.3 kpc), albeit with a poor confidence parameter.  In N65, \citet{Andersonbania2009} resolved the kinematic distance uncertainty toward the H{\small II} region located in N65 using HI emission/absorption, and determined the source was at the near distance.  They determined a distance of 3.6 kpc toward N65; this distance is consistent with the value of 3.6 kpc used by \citet{Deharveng2010} and 3.4 kpc used by \citet{brogan2011} and \citet{cyganowski2011}.  \citet{reid2016} measured parallax toward one of the methanol masers detected by \citet{cyganowski2009}.  If we assume that N65 is located with this maser, than parallax gives a distance of ~2.3 kpc to N65, consistent with a near distance. Placing N62 and N65 at the near distance provides more typical bubble sizes of  $\sim$6 and 10 pc, respectively (\citet{Deharveng2010}, assuming a distance of 3.7 kpc and 3.6 kpc).  Far distances of 10.5 kpc would put the sizes of N62 and N65 closer to 18 pc and 30 pc, significantly larger than the typical bubble size of $<5$ pc reported in \citet{Simpson2012}. 

The 2.68 kpc distance returned by the \citet{reid2016} Bayesian distance calculator for N90 is significantly different than the distance of $\sim 6$ pc reported for this bubble in \citet{Deharveng2010} and \citet{Anderson2012}.  However, as we note in Section \ref{sec:N90source}, the velocity of the emission we detected toward this source is different than the velocity used in these studies, leading to the difference in the derived distances.  

\section{Observations}
\label{sec:observations}

Our observations were conducted 2015 March through 2015 May using the Versatile GBT Astronomical Spectrometer (VEGAS) instrument.  The VEGAS instrument consists of eight independent spectrometers, each with up to eight spectral windows, so that up to $64$ separate spectral windows can be observed simultaneously.  We observed $21$ lines using the Q-band receiver to verify and expand upon the results of \citet{Watson2016}.  A full list of the lines observed and their rest frequencies is provided in Table \ref{tab:lines}.  Coverage of our line sample required two configurations of the VEGAS instrument. In addition to CS (1-0), we observed its optically thin isotope, C$^{34}$S (1-0), to test the infall model proposed in \citet{Watson2016}.  We also observed additional dense gas tracers (e.g. H$_2$CO, HC$_3$N(5-4)) and transitions often associated with outflows and star formation (e.g. SiO (1-0), CH$_3$OH (1-0), and radio recombination lines).  

\begin{figure}
\begin{center}
\includegraphics[width=0.7\columnwidth]{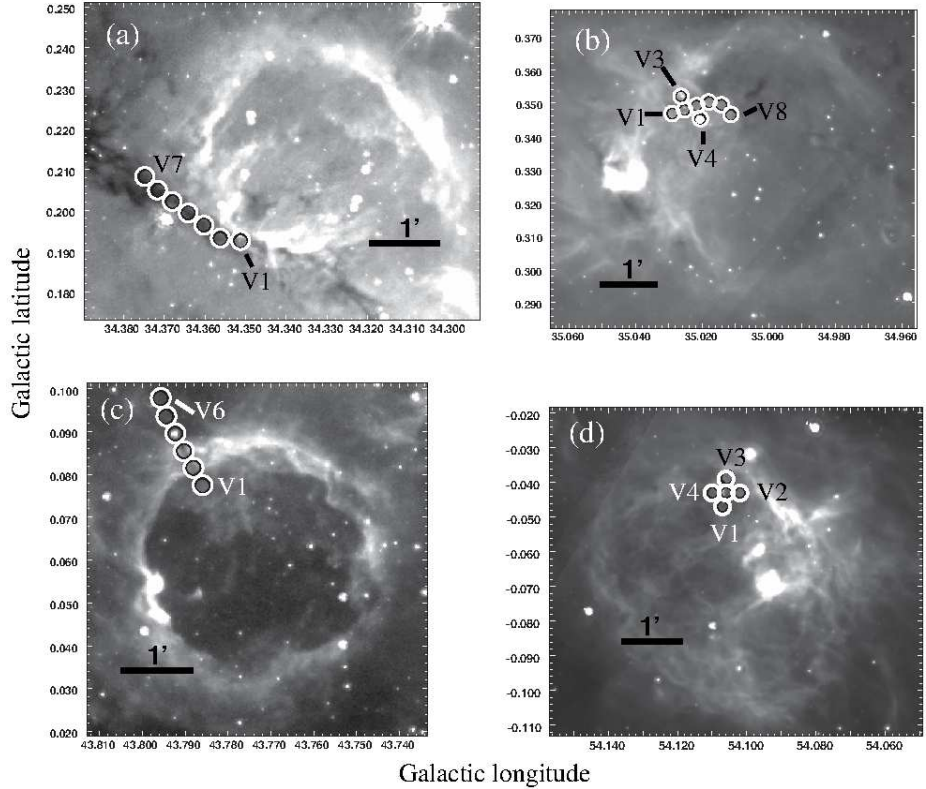}
\caption{ Spitzer Space Telescope GLIMPSE 8 $\mu$m images of source (a) N62, (b) N65, (c) N90, and (d) N117. Circles indicate the GBT pointings with $16''$ FWHM beam size. Select pointings are labeled.  The black line in each image indicates an angular size of 1 \arcmin. }

\label{fig:bubbles}
\end{center}
\end{figure}

\begin{figure}
\begin{center}
\includegraphics[width=0.7\columnwidth]{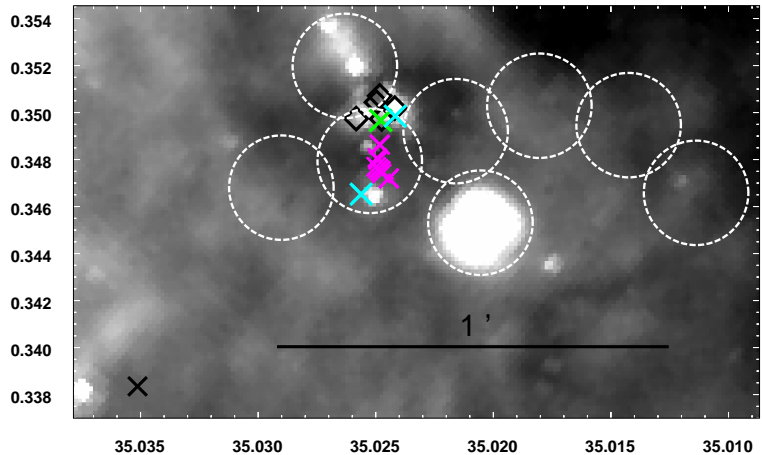}
\caption{Spitzer Space Telescope GLIMPSE 8 $\mu$m image of source N65, zoomed in on the pointings shown in Fig. \ref{fig:bubbles}b  and showing the location of other notable sources near our pointings. White, dashed circles indicate the GBT pointings with $16''$ FWHM beam size.  The line in the lower right indicates an angular size of 1 \arcmin. The black X marks the location of the UC  H {\small II} region detected by \citet{kurtz2005}, the green X indicates the location of the Class II methanol maser reported by \citet{cyganowski2009}, the magenta X's indicate the location of the chain of Class 1 44 GHz methanol masers reported in \citet{cyganowski2009}, the black diamonds indicate the location of the five compact continuum sources reported in \citet{cyganowski2011}, and the cyan X's indicate the location of the 25 GHz methanol masers detected by \citet{brogan2011}.}

\label{fig:N65fig}
\end{center}
\end{figure}

A total of $26$ pointings were conducted. Each pointing was observed for two four-minute integrations in frequency-switching mode.  The parameters for GBT setup and observing and the calibration sources are listed in Table \ref{tab:obs_params}.  We observed in dual-beam mode. One beam was aimed at our target, the second beam position was determined by the rotation angle of the receiver and not consistent between observations.  To maximize observing efficiency, flux calibration was only performed for the first, targeted beam. The pointings shown in Fig. \ref{fig:bubbles} are only the first, calibrated beam, and the results we present are solely from this beam with the exception of discussion in Section \ref{sec:kinematics}. The data were calibrated and analyzed using GBTIDL \citep{GBTIDLref}.  

We assumed a linear baseline near the emission lines during baseline calibration. Finally, we averaged data from both polarizations and all scans for each pointing. Noise was determined by taking the standard deviation of 200 channels of data near but offset from the emission lines.  Typical rms noise in the calibrated data was $\sim 0.1$ K, except in pointing N65 V6, which had a typical rms noise value of $\sim 0.4$ K. Typical system temperatures were between $70$ K and $115$ K.  We estimate our uncertainty due to flux calibration to be $\sim 20\%$.  Pointing accuracy is estimated from the GBT Observer's Manual, which states that tracking error is $\le 20\%$ of the beam size. 

In the analysis presented in Section \ref{sec:abundances}, we use data from the Hi-Gal Project \citep{Molinari2010}.  Hi-Gal is a far-infrared (FIR) survey of the Galactic plane conducted with the Herschel Space Telescope at wavelengths between 60 and 600 $\mu$m. Fully calibrated, level 2.5 data products were downloaded from the Herschel Science Archive. 

\begin{table}
\caption{Source Names and Positions}
\label{tab:pointings}
\begin{tabular}{lcc}

\hline
Name    &l $^\circ$  &b $^\circ$\\
\hline
N62 V1 &  34.351 & 0.192\\
N62 V2 &  34.356 & 0.193\\
N62 V3 &  34.360 & 0.196\\
N62 V4 &  34.364 & 0.199\\
N62 V5 &  34.368 & 0.202\\
N62 V6 &  34.371 & 0.204\\
N62 V7 &  34.374 & 0.208\\
\hline
N65 V1 & 35.029 & 0.346 \\
N65 V2 & 35.025 & 0.347 \\
N65 V3 & 35.026 & 0.351\\
N65 V4 & 35.020 & 0.345\\
N65 V5 & 35.021 & 0.349\\
N65 V6 & 35.018 & 0.350\\
N65 V7 & 35.014 & 0.349\\
N65 V8 & 35.011 & 0.346\\
\hline
N90 V1 & 43.795 & 0.097\\
N90 V2 & 43.794 & 0.093\\
N90 V3 & 43.792 & 0.089\\
N90 V4 & 43.790 & 0.085\\
N90 V5 & 43.788 & 0.081\\
N90 V6 & 43.785 & 0.077\\
\hline
N117 V1 & 54.107 & -0.047\\
N117 V2 & 54.102 & -0.043\\
N117 V3 & 54.106 & -0.039\\
N117 V4 & 54.110 & -0.043\\
N117 V5 & 54.106 & -0.043\\

\end{tabular}
\end{table}

\begin{table}
\caption{Molecular lines observed.}
\label{tab:lines}
\begin{tabular}{lcc}
\hline
Molecule & Transition & Rest Frequency (GHz) \\
\hline
C$_3$N & 4-3 & 39.5714 \\
H 68 $\beta$ & & 40.05288 \\
CCCS & 7$_0$-6$_0$ & 40.4650 \\
H 54 $\alpha$ & & 40.6305 \\
He 54 $\alpha$ & & 40.64706 \\
SiO (v=3) & 1-0 & 42.5193 (*)\\
HCS+ & 1-0 & 42.6742 \\
SiO (v=2) & 1-0 & 42.8204 (*) \\
$^{29}$SiO (v=0) & 1-0 & 42.8798 (*) \\
SiO (v=0) & 1-0 & 43.4238 \\ 
H 66 $\alpha$ & & 43.74896 \\
He 66 $\beta$ & & 43.76678 \\
 HC$_3$N & 5-4 & 45.4903  \\
 H 52 $\alpha$ &   & 45.45372 \\
 He 52 $\alpha$ &   & 45.47225  \\
 H 65 $\beta$ &   & 45.76845 \\
 C$^{34}$S & 1-0  & 48.2069 \\
H$_2$CO & 4$_{1,3}$-4$_{1,4}$  & 48.28455 \\
 CH$_3$OH & 1(0,0)$^+$-0(0,0)$^+$ &  48.37246 \\
 OCS & 4-3 & 48.6516 \\
 CS & 1-0 & 48.99098 \\
\end{tabular}

\textbf{Notes.} Molecular line rest frequencies were obtained from the JPL Line Catalog \citep{Pickett1998} with the exception of sources marked with (*), which were taken from the CDMS Line List \citep{Muller2001, Muller2005}. Radio recombination lines were obtained from \citet{lilleypalmer1968}.
\end{table}

\begin{table}
\caption{Observing Parameters}
\label{tab:obs_params}
\begin{tabular}{l c}
\hline
   Bandwidth & 23.44 MHz\\ 
   Channel width & 5.722 KHz\\ 
   Frequency switching shift & 8.0 MHz\\ 
   Angular resolution & 16 \arcsec \\
   Pointing calibration & 1850-0001 \\
    & 1851+005 \\

    & 1751+0939 \\
    & 2015+3710 \\
   Flux calibration & NGC7027 \\
\hline
\end{tabular}
\end{table}

\begin{sidewaystable}

\caption{Gaussian Fitting Parameters.}

\hspace{-1in}\resizebox{1.1\textwidth}{!}{
\begin{tabular}{lcccccccccccccccc}
\hline
 & & HC$_3$N & & & C$^{34}$S & & & CH$_3$OH & & & CS & & & & SiO & \\
Source & $T$\textsubscript{mb} & $V$\textsubscript{LSR} & FWHM & $T$\textsubscript{mb} & $V$\textsubscript{LSR} & FWHM & $T$\textsubscript{mb} & $V$\textsubscript{LSR} & FWHM & $T$\textsubscript{mb} & $V$\textsubscript{LSR} & FWHM & First Moment & $T$\textsubscript{mb} & $V$\textsubscript{LSR} & FWHM \\
 & (K) & (km s$^{-1}$) & (km s$^{-1}$) & (K) & (km s$^{-1}$) & (km s$^{-1}$) & (K) & (km s$^{-1}$) & (km s$^{-1}$) & (K) & (km s$^{-1}$) & (km s$^{-1}$) & (km s$^{-1}$) & (K) & (km s$^{-1}$) & (km s$^{-1}$) \\
\hline
N62 V1 & $\cdots$ & $\cdots$ & $\cdots$ & $\cdots$ & $\cdots$ & $\cdots$ & $\cdots$ & $\cdots$ & $\cdots$ & $2.21 \pm 0.04$ & $56.59 \pm 0.02$ & $1.20 \pm 0.04$ & $56.99$ & $\cdots$ & $\cdots$ & $\cdots$ \\
& $\cdots$ & $\cdots$ & $\cdots$ & $\cdots$ & $\cdots$ & $\cdots$ & $\cdots$ & $\cdots$ & $\cdots$ & $1.14 \pm 0.06$ & $57.75 \pm 0.03$ & $0.84 \pm 0.06$ & $\cdots$ & $\cdots$  & $\cdots$  & $\cdots$ \\
N62 V2 & $0.30 \pm 0.02$ & $56.43 \pm 0.04$ & $1.02 \pm 0.08$ & $0.47 \pm 0.03$ & $56.42 \pm 0.03$ & $0.98 \pm 0.07$ & $\cdots$ & $\cdots$ & $\cdots$ & $3.69 \pm 0.04$ & $56.72 \pm 0.01$ & $1.53 \pm 0.02$ & $56.78$ & $\cdots$ & $\cdots$ & $\cdots$ \\
N62 V3 & $0.44 \pm 0.02$ & $56.87 \pm 0.03$ & $1.26 \pm 0.06$ & $0.41 \pm 0.03$ & $56.81 \pm 0.04$ & $1.24 \pm 0.09$ & $\cdots$ & $\cdots$ & $\cdots$ & $3.66 \pm 0.04$ & $56.88 \pm 0.02$ & $1.28 \pm 0.32$ & $57.06$ & $\cdots$ & $\cdots$ & $\cdots$ \\
& $\cdots$ & $\cdots$ & $\cdots$ & $\cdots$ & $\cdots$ & $\cdots$ & $\cdots$ & $\cdots$ & $\cdots$ & $1.09 \pm 0.08$ & $57.98 \pm 0.04$ & $0.93 \pm 0.07$ & $\cdots$ & $\cdots$  & $\cdots$  & $\cdots$ \\
N62 V4 & $0.78 \pm 0.02$ & $56.96 \pm 0.01$ & $0.94 \pm 0.03$ & $0.47 \pm 0.04$ & $56.84 \pm 0.02$ & $0.59 \pm 0.05$ & $0.40 \pm 0.03$ & $56.96 \pm 0.04$ & $0.97 \pm 0.09$ & $3.61 \pm 0.04$ & $57.03 \pm 0.01$ & $1.14 \pm 0.03$ & $56.96$ & $\cdots$ & $\cdots$ & $\cdots$ \\
& $\cdots$ & $\cdots$ & $\cdots$ & $\cdots$ & $\cdots$ & $\cdots$ & $\cdots$ & $\cdots$ & $\cdots$ & $0.83 \pm 0.07$ & $58.11 \pm 0.03$ & $0.70 \pm 0.07$ & $\cdots$ & $\cdots$  & $\cdots$  & $\cdots$ \\
N62 V5 & $0.82 \pm 0.02$ & $57.16 \pm 0.01$ & $1.16 \pm 0.03$ & $\cdots$ & $\cdots$ & $\cdots$ & $0.44 \pm 0.03$ & $57.07 \pm 0.03$ & $1.07 \pm 0.08$ & $3.68 \pm 0.05$ & $57.04 \pm 0.02$ & $1.16 \pm 0.03$ & $57.25$ & $\cdots$ & $\cdots$ & $\cdots$ \\
& $\cdots$ & $\cdots$ & $\cdots$ & $\cdots$ & $\cdots$ & $\cdots$ & $\cdots$ & $\cdots$ & $\cdots$ & $1.55 \pm 0.06$ & $58.25 \pm 0.04$ & $1.02 \pm 0.06$ & $\cdots$ & $\cdots$ & $\cdots$ & $\cdots$ \\
N62 V6 & $0.63 \pm 0.02$ & $57.26 \pm 0.02$ & $1.07 \pm 0.04$ & $0.46 \pm 0.03$ & $57.13 \pm 0.04$ & $1.28 \pm 0.08$ & $0.46 \pm 0.03$ & $57.16 \pm 0.03$ & $1.29 \pm 0.08$ & $3.21 \pm 0.04$ & $57.24 \pm 0.02$ & $1.54 \pm 0.04$ & $57.23$ & $\cdots$ & $\cdots$ & $\cdots$ \\
& $\cdots$ & $\cdots$ & $\cdots$ & $\cdots$ & $\cdots$ & $\cdots$ & $\cdots$ & $\cdots$ & $\cdots$ & $1.26 \pm 0.07$ & $58.63 \pm 0.03$ & $0.81 \pm 0.06$ & $\cdots$ & $\cdots$ & $\cdots$ & $\cdots$ \\
N62 V7 & $0.28 \pm 0.01$ & $57.4 \pm 0.03$ & $1.09 \pm 0.07$ & $0.56 \pm 0.03$ & $57.21 \pm 0.02$ & $0.92 \pm 0.05$ & $0.48 \pm 0.02$ & $57.37 \pm 0.03$ & $1.16 \pm 0.07$ & $3.15 \pm 0.04$ & $57.33 \pm 0.01$ & $1.32 \pm 0.03$ & $57.40$ & $\cdots$ & $\cdots$ & $\cdots$ \\
& $\cdots$ & $\cdots$ & $\cdots$ & $\cdots$ & $\cdots$ & $\cdots$ & $\cdots$ & $\cdots$ & $\cdots$ & $1.00 \pm 0.06$ & $58.63 \pm 0.03$ & $0.81 \pm 0.06$ & $\cdots$ & $\cdots$ & $\cdots$ & $\cdots$ \\
\hline
N65 V1 & $1.79 \pm 0.02$ & $52.51 \pm 0.01$ & $1.86 \pm 0.02$ & $1.20 \pm 0.02$ & $52.15 \pm 0.02$ & $1.66 \pm 0.04$ & $0.41 \pm 0.02$ & $52.70 \pm 0.07$ & $3.22 \pm 0.17$ & $4.07 \pm 0.39$ & $51.79 \pm 0.05$ & $1.21 \pm 0.07$ & $52.47$ & $0.35 \pm 0.02$ & $52.42 \pm 0.05$ & $2.50 \pm 0.13$\\
& $\cdots$ & $\cdots$ & $\cdots$ & $\cdots$ & $\cdots$ & $\cdots$ & $\cdots$ & $\cdots$ & $\cdots$ &  $6.44 \pm 0.20$ & $52.94 \pm 0.05$ & $1.50 \pm 0.07$ & $\cdots$ & $\cdots$ & $\cdots$ & $\cdots$ \\
N65 V2 & $2.33 \pm 0.01$ & $52.68 \pm 0.01$ & $3.51 \pm 0.02$ & $1.95 \pm 0.02$ & $52.19 \pm 0.02$ & $3.31 \pm 0.04$ & $0.65 \pm 0.02$ & $52.79 \pm 0.05$ & $4.16 \pm 0.13$ & $8.74 \pm 0.16 $ & $51.55 \pm 0.04$ & $1.83 \pm 0.13 $ & $52.89$ & $0.32 \pm 0.01$ & $52.55 \pm 0.07$ & $3.77 \pm 0.18$ \\
& $\cdots$ & $\cdots$ & $\cdots$ & $\cdots$ & $\cdots$ & $\cdots$ & $\cdots$ & $\cdots$ & $\cdots$ & $11.41 \pm 0.24$ & $53.42 \pm 0.03$ & $1.73 \pm 0.07$ & $\cdots$ & $\cdots$ & $\cdots$ & $\cdots$ \\
N65 V3 & $1.24 \pm 0.01$ & $52.47 \pm 0.02$ & $3.19 \pm 0.04$ & $1.12 \pm 0.02$ & $52.44 \pm 0.02$ & $2.45 \pm 0.04$ & $1.10 \pm 0.03$ & $52.44 \pm 0.03$ & $2.38 \pm 0.07$ & $7.98 \pm 0.03$ & $52.79 \pm 0.01$ & $3.07 \pm 0.03$ & $52.89$ & $0.36 \pm 0.01$ & $52.40 \pm 0.05$ & $2.45 \pm 0.12$ \\
N65 V4 & $0.24 \pm 0.01$ & $53.91 \pm 0.08$ & $2.91 \pm 0.18$ & $0.42 \pm 0.02$ & $53.63 \pm 0.04$ & $1.90 \pm 0.10$ & $0.62 \pm 0.02$ & $53.92 \pm 0.05$ & $2.87 \pm 0.11$ & $0.83 \pm 0.03$ & $51.35 \pm 0.05$ & $1.45 \pm 0.15$ & $53.69$ & $\cdots$ & $\cdots$ & $\cdots$ \\
& $\cdots$ & $\cdots$ & $\cdots$ & $\cdots$ & $\cdots$ & $\cdots$ & $\cdots$ & $\cdots$ & $\cdots$ & $4.88 \pm 0.03$ & $53.90 \pm 0.01$ & $1.92 \pm 0.02$ & $\cdots$ & $\cdots$ & $\cdots$  & $\cdots$ \\
N65 V5 & $1.10 \pm 0.02$ & $54.05 \pm 0.02$ & $3.22 \pm 0.05$ & $1.05 \pm 0.02$ & $53.76 \pm 0.02$ & $2.83 \pm 0.05$ & $0.82 \pm 0.02$ & $53.78 \pm 0.05$ & $3.83 \pm 0.11$ & $9.44 \pm 0.05$ & $53.93 \pm 0.02$ & $2.42 \pm 0.05$ & $54.22$ & $\cdots$ & $\cdots$ & $\cdots$ \\
& $\cdots$ & $\cdots$ & $\cdots$ & $\cdots$ & $\cdots$ & $\cdots$ & $\cdots$ & $\cdots$ & $\cdots$ & $3.01 \pm 0.22 $ & $55.59 \pm 0.02$ & $1.36 \pm 0.11 $ & $\cdots$ & $\cdots$ & $\cdots$ & $\cdots$ \\
N65 V6 & $1.70 \pm 0.05$ & $53.92 \pm 0.03$ & $1.96 \pm 0.07$ & $2.17 \pm 0.07$ & $53.78 \pm 0.03$ & $2.04 \pm 0.08$ & $4.12 \pm 0.08$ & $54.13 \pm 0.02$ & $2.46 \pm 0.06$ & $12.23 \pm 0.09$ & $54.11 \pm 0.01$ & $2.68 \pm 0.06$ & $54.06$ & $\cdots$ & $\cdots$ & $\cdots$ \\

N65 V7 & $0.48 \pm 0.01$ & $52.98 \pm 0.03$ & $2.46 \pm 0.08$ & $0.80 \pm 0.02$ & $52.80 \pm 0.03$ & $2.30 \pm 0.06$ & $0.89 \pm 0.02$ & $53.20 \pm 0.03$ & $2.96 \pm 0.07$ & $3.87 \pm 0.26 $ & $52.40 \pm 0.07$ & $1.68 \pm 0.10 $ & $53.14$ & $\cdots$ & $\cdots$ & $\cdots$ \\
& $\cdots$ & $\cdots$ & $\cdots$ & $\cdots$ & $\cdots$ & $\cdots$ & $\cdots$ & $\cdots$ & $\cdots$ & $3.67 \pm 0.22 $ & $53.92 \pm 0.08$ & $1.77 \pm 0.12 $ & $\cdots$ & $\cdots$  & $\cdots$  & $\cdots$ \\

N65 V8 & $0.66 \pm 0.02$ & $52.08 \pm 0.02$ & $1.46 \pm 0.04$ & $1.17 \pm 0.03$ & $51.97 \pm 0.02$ & $1.43 \pm 0.04$ & $0.73 \pm 0.02$ & $52.10 \pm 0.04$ & $2.23 \pm 0.09$ & $5.33 \pm 0.04$ & $52.25 \pm 0.02$ & $2.05 \pm 0.03$ & $52.61$ & $\cdots$ & $\cdots$ & $\cdots$ \\
& $\cdots$ & $\cdots$ & $\cdots$ & $\cdots$ & $\cdots$ & $\cdots$ & $\cdots$ & $\cdots$ & $\cdots$ & $1.76 \pm 0.07$ & $54.06 \pm 0.03$ & $1.40 \pm 0.06$ & $\cdots$ & $\cdots$ & $\cdots$ & $\cdots$ \\
\hline
N90 V3 & $\cdots$ & $\cdots$ & $\cdots$ & $\cdots$ & $\cdots$ & $\cdots$ & $\cdots$ & $\cdots$ & $\cdots$ & $1.15 \pm 0.04$ & $35.76 \pm 0.02$ & $1.03 \pm 0.04$ & $34.94$ & $\cdots$ & $\cdots$ & $\cdots$ \\
N90 V4 & $\cdots$ & $\cdots$ & $\cdots$ & $\cdots$ & $\cdots$ & $\cdots$ & $\cdots$ & $\cdots$ & $\cdots$ & $1.46 \pm 0.04$ & $35.47 \pm 0.01$ & $0.91 \pm 0.03$ & $34.79$ & $\cdots$ & $\cdots$ & $\cdots$ \\
N90 V5 & $\cdots$ & $\cdots$ & $\cdots$ & $\cdots$ & $\cdots$ & $\cdots$ & $\cdots$ & $\cdots$ & $\cdots$ & $1.09 \pm 0.05$ & $35.45 \pm 0.01$ & $0.61 \pm 0.03$ & $33.94$ & $\cdots$ & $\cdots$ & $\cdots$ \\
N90 V6 & $\cdots$ & $\cdots$ & $\cdots$ & $\cdots$ & $\cdots$ & $\cdots$ & $\cdots$ & $\cdots$ & $\cdots$ & $0.61 \pm 0.04$ & $35.49 \pm 0.02$ & $0.58 \pm 0.05$ & $34.46$ & $\cdots$ & $\cdots$ & $\cdots$ \\
\hline
N117 V1 & $0.22 \pm 0.01$ & $38.98 \pm 0.06$ & $1.72 \pm 0.13$ & $\cdots$ & $\cdots$ & $\cdots$ & $\cdots$ & $\cdots$ & $\cdots$ & $3.43 \pm 0.03$ & $38.82 \pm 0.02$ & $2.10 \pm 0.04$ & $39.40$ & $\cdots$ & $\cdots$ & $\cdots$ \\
& $\cdots$ & $\cdots$ & $\cdots$ & $\cdots$ & $\cdots$ & $\cdots$ & $\cdots$ & $\cdots$ & $\cdots$ & $2.40 \pm 0.05$ & $40.80 \pm 0.02$ & $1.38 \pm 0.04$ & $\cdots$& $\cdots$ & $\cdots$ & $\cdots$ \\
N117 V2 & $\cdots$ & $\cdots$ & $\cdots$ & $\cdots$ & $\cdots$ & $\cdots$ & $\cdots$ & $\cdots$ & $\cdots$ & $1.54 \pm 0.02$ & $38.69 \pm 0.03$ & $2.20 \pm 0.07$ & $39.86$ & $\cdots$ & $\cdots$ & $\cdots$ \\
& $\cdots$ & $\cdots$ & $\cdots$ & $\cdots$ & $\cdots$ & $\cdots$ & $\cdots$ & $\cdots$ & $\cdots$ & $1.85 \pm 0.03$ & $41.00 \pm 0.02$ & $1.73 \pm 0.05$ & $\cdots$ & $\cdots$ & $\cdots$ & $\cdots$ \\
N117 V3 & $\cdots$ & $\cdots$ & $\cdots$ & $\cdots$ & $\cdots$ & $\cdots$ & $\cdots$ & $\cdots$ & $\cdots$ & $1.92 \pm 0.03$ & $38.96 \pm 0.04$ & $2.54 \pm 0.08$ & $39.83$ & $\cdots$ & $\cdots$ & $\cdots$ \\
& $\cdots$ & $\cdots$ & $\cdots$ & $\cdots$ & $\cdots$ & $\cdots$ & $\cdots$ & $\cdots$ & $\cdots$ & $1.68 \pm 0.06$ & $41.18 \pm 0.03$ & $1.40 \pm 0.06$ & $\cdots$ & $\cdots$ & $\cdots$ & $\cdots$ \\
N117 V4 & $\cdots$ & $\cdots$ & $\cdots$ & $\cdots$ & $\cdots$ & $\cdots$ & $\cdots$ & $\cdots$ & $\cdots$ & $2.84 \pm 0.03$ & $38.78 \pm 0.04$ & $2.57 \pm 0.07$ & $39.02$ & $\cdots$ & $\cdots$ & $\cdots$ \\
& $\cdots$ & $\cdots$ & $\cdots$ & $\cdots$ & $\cdots$ & $\cdots$ & $\cdots$ & $\cdots$ & $\cdots$ & $0.85 \pm 0.07$ & $40.98 \pm 0.09$ & $1.72 \pm 0.15$ & $\cdots$ & $\cdots$ & $\cdots$ & $\cdots$ \\
N117 V5 & $0.33 \pm 0.02$ & $38.77 \pm 0.06$ & $1.75 \pm 0.13$ & $0.38 \pm 0.03$ & $38.79 \pm 0.08$ & $1.88 \pm 0.19$ & $\cdots$ & $\cdots$ & $\cdots$ & $3.28 \pm 0.04$ & $38.77 \pm 0.03$ & $2.35 \pm 0.06$ & $39.37$ & $\cdots$ & $\cdots$ & $\cdots$ \\
& $\cdots$ & $\cdots$ & $\cdots$ & $\cdots$ & $\cdots$ & $\cdots$ & $\cdots$ & $\cdots$ & $\cdots$ & $1.79 \pm 0.06$ & $40.99 \pm 0.04$ & $1.70 \pm 0.07$ & $\cdots$ & $\cdots$ & $\cdots$ & $\cdots$ \\
\end{tabular}}

\textbf{Notes.}  Only detections with T${_{mb}}^*$ greater than $3 \sigma$ are reported; noise values were $\sim 0.1$K, except for N65 V6, which has noise of about $\sim 0.4$K. First moments are reported for CS emission.

\label{LineData}

\end{sidewaystable}

\section{Results and Analysis}
\label{sec:results}

Lines were fit with the ``fitgauss'' procedure from GBTIDL, which returns the peak T$_{mb}$, central LSR velocity, and FWHM of the line. The lines observed toward each pointing are reported in Table \ref{LineData}, along with the Gaussian fit parameters and the fitting uncertainty. Lines with non-Gaussian, double-peak profiles were simultaneously fit with two Gaussians, and the results of both fits are reported in Table \ref{LineData} in consecutive rows. Each fit was visually inspected to ensure accuracy. Only detections with peak T$_{mb} > 3 \sigma$ are reported. Because of the complicated line structure in some of the CS (1-0) lines, the first moment (i.e. the sum of  the emission over the full velocity range of the emission line) of the CS lines was also calculated using GBTIDL task ``gmoment'' for use in our kinematics analysis. The first moment calculation is used in our analysis of the gas kinematics in Section \ref{sec:kinematics}.  Lines listed in Table \ref{tab:lines} but not Table \ref{LineData} were not detected in any of our observations above the $3\sigma$ noise cut-off. 

The lines we detect are useful probes of dense gas and star formation.  Due to its large critical density, CS (1-0) is a dense gas tracer ($\sim 10^4$ at temperatures of 10-20 K, \citep{shirley2015}).  The C$^{34}$S (1-0) transition provides an optically thin line to compare to the CS (1-0) emission, to check whether asymmetric line profiles are the result of self absorption or a different phenomenon. HC$_3$N (5-4) traces even higher density gas, with a critical density of $\sim 5$x$10^4$, and observations by \citet{meier2011} provide some evidence that  HC$_3$N (5-4) generally traces quiescent (rather than actively star forming) gas. SiO emission traces high temperature and/or shocks, potentially from outflows \citep{Bachiller96,martin97}.  Thermal CH$_3$OH is a tracer of hot, dense molecular cores and/or shocked gas \citep{mehringer97}.

We note that two of our pointings, N62 V1 and N90 V3, align exactly with CS (1-0) observations reported in \citet{Watson2016} but the fluxes we report are lower than those in \citet{Watson2016} by a larger amount than can be accounted for with flux calibration uncertainty ($\sim 35$\% for N62 V1 and $\sim 80$\% for N90 V3).  One possible explanation for this discrepancy is pointing uncertainty toward diffuse but inhomogeneous sources, such that a $20$\% offset in positioning could lead to significant flux differences. Unfortunately, the best way to confirm this interpretation would be to obtain a high-resolution map of the region, which is outside the scope of the present study.  To determine if such a large decrease is plausibly caused by actual changes in the source, we make the following estimates. At the distances reported above, and assuming a sound speed of $0.2$ km s$^{-1}$, the sound crossing time for the entire GBT beam is $\sim 10^6$ years. A large object changing so dramatically in brightness over a period much smaller than the sound crossing time seems implausible. If the object were smaller than the GBT beam, however, quick changes in brightness would be more reasonable. The sound crossing time and two years’ time interval between GBT observations implies a size of $0.08$ AU. It is implausible that such a small object could be so bright in a molecular line. The only exception to this argument would be a maser, but we are aware of no reported masing in this molecular line (CS at $49$ GHz). Thus, we find it implausible that the observed flux differences for N62 V1 and N90 V3 represent changes in the source flux.

\subsection{Column Densities and Abundances} 
\label{sec:abundances}
Column densities for HC$_3$N, C$^{34}$S, CH$_3$OH, CS, and SiO were calculated using the following relation \citep{Miettinen2012}:

\begin{equation}
N = \dfrac{3k_B\epsilon_0}{2\pi^2}\dfrac{1}{\nu\mu^2_{el}S}\dfrac{Z_{rot}(T_{ex})}{g_Kg_I}\dfrac{e^{E_u/k_B T_{ex}}}{1 - \frac{F(T_{bg})}{F(T_{ex})}}\int T_{MB}dv
\end{equation}

where

\begin{equation*}
F(T) = \dfrac{1}{e^{h\nu/k_BT} - 1}.
\end{equation*}

\noindent In these relations $k_B$ is the Boltzmann constant, $\epsilon_0$ is the vacuum permittivity, $\nu$ is the rest frequency of the molecule, $\mu_{el}^2S$ is the electric dipole moment line strength, $Z_{rot}$ is the partition function, $T_{ex}$ and $T_{bg}$ are the excitation and background temperatures, $g_K$ is the K-level degeneracy, $g_I$ is the reduced nuclear spin degeneracy, $E_u$ is the upper energy state, $h$ is Planck's constant, and the final term in the equation is the integrated emission based on the Gaussian fit parameters given in Table \ref{LineData}. We assume optically thin emission, local thermal equilibrium, a background temperature of $2.725$ K from the cosmic microwave background, and an excitation temperature value of $20$ K.  The assumed excitation temperature is based on typical values found when we fit modified black-body curves to Hi-GAL dust emission from the sources (see discussion below), and is a temperature consistent with typical ISM values in star forming environments \citep{Zinnecker2007}. 

The values used for the other parameters are presented in Table \ref{tab:col_dens_params}. Where double-Gaussian profiles were fit to the CS emission, the column density contribution from both lines were added to determine the total CS column density.  The column densities are given in Table \ref{tab:columndensity}.  

The excitation temperature estimate is the largest source of uncertainty in the assumptions used to determine column density. Although the modified blackbody fitting of the broadband FIR emission has relatively small statistical errors, the dust and gas emission probably do not sample the material with equal weighting.  When we varied the excitation temperature over a range of $\pm 5$ K, from $15$ K to $25$ K, the column densities of HC$_3$N, CH$_3$OH, C$^{34}$S, CS, and SiO varied by $\sim\ 15$\%, $20$ \%, $30$\%, $20$\%, and $20$\% respectively. Varying the excitation temperature over a range of $\pm 10$ K increased these uncertainties by roughly an additional factor of three. 

We calculated abundances of HC$_3$N, CH$_3$OH, C$^{34}$S, CS, and SiO by combining the calculated column densities with the total gas column density.  We calculated the total column density using FIR dust emission imaged as part of the Hi-GAL survey and applying the analysis described in detail in \citet{Watson2016}.  In summary, we measured the integrated emission in all five Hi-GAL survey bands in regions coincident with the $16''$ GBT beam, centered at each pointing in Table \ref{tab:pointings}. The emission was then modeled as a modified blackbody ($\beta$ = 2) and the total column density was calculated following \citet{Miettinen_2010}. The integrated emission in each Hi-GAL survey band, the total column density, and the molecular abundances are listed in Table \ref{tab:abundance}. As discussed in \citet{Watson2016}, these column densities and abundances should be interpreted cautiously because the FIR emission is usually extended in these regions and the FIR emission and molecular line emissions probably do not sample the same regions equally. A $20$\% change in FIR flux had a typical change of $4$ K and $20$\% in total column density.

We calculated the relative abundance of [C$^{34}$S]/[CS] to be $\sim 0.1$.  This value is higher than the typically assumed value of $0.045$ \citep{wilsonrood94}, but since CS is likely to be optically thick in most of our detections, the CS column densities and abundances should be treated as lower limits, making our abundance ratio an upper limit.  

\begin{table}
\caption{Values of parameters used to calculate column density.}
\label{tab:col_dens_params}
\begin{tabular}{ccccccc}
\hline
& CS & C$^{34}$S & HC$_3$N & CH$_3$OH & SiO & Reference\\
\hline
$g_Kg_I$ & $1$ & $1$ & $1$ & $2$ & $1$ & 1 \\
$\mu^2_{el}S$ & $3.82$ Debye$^2$ & $3.8$ Debye$^2$ & $69.6$ Debye$^2$ & $0.8$ Debye$^2$ & $9.6$ Debye$^2$ & 2 \\
$Z_{rot}(T_{ex})$ & $0.850\ T_{ex} + 0.347$ & $0.864\ T_{ex} + 0.348$ & $13.7\ T_{ex} + 1.004$ & $5.25\ T_{ex} - 29.7$ & $0.959\ T_{ex} + 0.346$ & 3\\
$\nu $ & $48.9909549$ GHz & $48.2069150$ GHz & $45.4903102$ GHz & $48.372460$ GHz & $43.423760$ GHz & 4\\
$E_u$ & $1.6342$ cm$^{-1}$ & $2.2950$ cm$^{-1}$ & $4.5522$ cm$^{-1}$ & $1.614$ cm$^{-1}$ & $1.45$ cm$^{-1}$ & 4\\
\end{tabular}

\medskip
\textbf{References.} (1) \citet{Turner1991}, (2) CDMS Catalog (\citet{Muller2001} and \citet{Muller2005}), (3) Linear interpolation from JPL line catalog \citep{Pickett1998}, (4) JPL line catalog \citep{Pickett1998}.
\end{table}

{\renewcommand{\arraystretch}{1.2} 
\begin{table} 
\caption{Column Densities} 
\label{tab:columndensity}
\begin{tabular}{lccccc} 
\hline
Source & N$_{HC_3N}$ & N$_{C^{34}S}$ & N$_{CH_3OH}$ & N$_{CS}$ & N$_{SiO}$ \\ 
 & ($\times 10^{12}$ cm$^{-2}$) & ($\times 10^{12}$ cm$^{-2}$) & ($\times 10^{12}$ cm$^{-2}$) & ($\times 10^{12}$ cm$^{-2}$) & ($\times 10^{12}$ cm$^{-2}$) \\ 
\hline 
N62 V1 & $\cdots$ & $\cdots$ & $\cdots$ & 74 & $\cdots$ \\ 
N62 V2 & 7 & 10 & $\cdots$ & 115 & $\cdots$ \\ 
N62 V3 & 13 & 112& $\cdots$ & 116 & $\cdots$ \\ 
N62 V4 & 17 & 6 & 82 & 96 & $\cdots$ \\ 
N62 V5 & 23 & $\cdots$ & 98 & 119 & $\cdots$ \\ 
N62 V6 & 16 & 13 & 126 & 121 & $\cdots$ \\ 
N62 V7 & 7 & 11 & 117 & 101 & $\cdots$ \\ 
\hline 
N65 V1 & 79 & 44 & 282 & 296 & 8 \\ 
N65 V2 & 194 & 142 & 572 & 728 & 12 \\ 
N65 V3 & 94 & 60 & 556 & 499 & 9 \\ 
N65 V4 & 17 & 18 & 379 & 215 & $\cdots$ \\ 
N65 V5 & 84 & 65 & 663 & 548 & $\cdots$ \\ 
N65 V6 & 79 & 98 & 2145 & 668 & $\cdots$ \\ 
N65 V7 & 28 & 41 & 558 & 265 & $\cdots$ \\ 
N65 V8 & 23 & 37 & 345 & 273 & $\cdots$ \\ 
\hline 
N117 V1 & 9 & $\cdots$ & $\cdots$ & 214 & $\cdots$ \\ 
N117 V2 & $\cdots$ & $\cdots$ & $\cdots$ & 135 & $\cdots$ \\ 
N117 V3 & $\cdots$ & $\cdots$ & $\cdots$ & 147 & $\cdots$ \\ 
N117 V4 & $\cdots$ & $\cdots$ & $\cdots$ & 179 & $\cdots$ \\ 
N117 V5 & 14 & 16 & $\cdots$ & 219 & $\cdots$ \\ 
\end{tabular} 
\end{table} 
}

{\renewcommand{\arraystretch}{1.2} 
\begin{sidewaystable}\centering 
\caption{Herschel Hi-Gal integrated emission in regions corresponding to GBT pointings, temperature obtained from modified blackbody fitting, calculated total column densities, and molecular abundances.} 
\label{tab:abundance} 
\begin{tabular}{l c c c c c c c c c c c c}
\hline
Source & Blue & Red & PSW & PMW & PLW & Temp. & N\textsubscript{tot} & HC$_3$N & C$^{34}$S & CH$_3$OH & CS & SiO \\ 
& $60-85 \mu$m & $130-210 \mu$m & $250 \mu$m & $350 \mu$m & $500 \mu$m & & & Abundance & Abundance & Abundance & Abundance & Abundance \\ 
& (Jy) & (Jy) & (Jy) & (Jy) & (Jy) & (K) & ($\times 10^{22}$ cm$^{-2}$) & ($\times 10^{-10}$) & ($\times 10^{-10}$) & ($\times 10^{-10}$) & ($\times 10^{-10}$) & ($\times 10^{-10}$) \\ 
\hline 
N62 V1 & 15 & 63 & 45 & 22 & 8 & 19 & 5 & $\cdots$ & $\cdots$ & $\cdots$ & 16 & $\cdots$ \\ 
N62 V2 & 13 & 60 & 44 & 22 & 8 & 18 & 5 & 2 & 2 & $\cdots$ & 23 & $\cdots$ \\ 
N62 V3 & 12 & 58 & 45 & 23 & 9 & 18 & 6 & 2 & 2 & $\cdots$ & 20 & $\cdots$ \\ 
N62 V4 & 10 & 56 & 45 & 24 & 9 & 17 & 6 & 3 & 1 & 13 & 15 & $\cdots$ \\ 
N62 V5 & 10 & 55 & 46 & 25 & 9 & 17 & 7 & 3 & $\cdots$ & 14 & 17 & $\cdots$ \\ 
N62 V6 & 8 & 52 & 45 & 24 & 10 & 16 & 8 & 2 & 2 & 17 & 16 & $\cdots$ \\ 
N62 V7 & 8 & 47 & 41 & 22 & 9 & 16 & 7 & 1 & 2 & 17 & 15 & $\cdots$ \\ 
\hline 
N65 V1 & 333 & 394 & 138 & 58 & 18 & 27 & 6 & 14 & 8 & 49 & 51 & 1 \\ 
N65 V2 & 1152 & 986 & 281 & 101 & 25 & 29 & 9 & 23 & 17 & 67 & 86 & 1 \\ 
N65 V3 & 1280 & 1124 & 349 & 119 & 27 & 29 & 10 & 9 & 6 & 56 & 50 & 1 \\ 
N65 V4 & 138 & 210 & 104 & 48 & 18 & 25 & 6 & 3 & 3 & 67 & 38 & $\cdots$ \\ 
N65 V5 & 813 & 747 & 247 & 92 & 25 & 28 & 8 & 10 & 8 & 82 & 68 & $\cdots$ \\ 
N65 V6 & 124 & 218 & 118 & 53 & 19 & 24 & 7 & 12 & 14 & 315 & 98 & $\cdots$ \\ 
N65 V7 & 33 & 92 & 58 & 28 & 12 & 21 & 5 & 6 & 9 & 119 & 56 & $\cdots$ \\ 
N65 V8 & 27 & 78 & 50 & 24 & 9 & 21 & 4 & 6 & 9 & 86 & 68 & $\cdots$ \\ 
\hline 
N117 V1 & 32 & 65 & 46 & 22 & 8 & 22 & 3 & 3 & $\cdots$ & $\cdots$ & 63 & $\cdots$ \\ 
N117 V2 & 34 & 52 & 38 & 18 & 6 & 23 & 2 & $\cdots$ & $\cdots$ & $\cdots$ & 59 & $\cdots$ \\ 
N117 V3 & 31 & 55 & 38 & 17 & 6 & 23 & 2 & $\cdots$ & $\cdots$ & $\cdots$ & 61 & $\cdots$ \\ 
N117 V4 & 26 & 54 & 41 & 19 & 7 & 22 & 3 & $\cdots$ & $\cdots$ & $\cdots$ & 60 & $\cdots$ \\ 
N117 V5 & 31 & 62 & 46 & 21 & 7 & 22 & 3 & 4 & 5 & $\cdots$ & 69 & $\cdots$ \\ 
\end{tabular} 
\end{sidewaystable} 
}

\subsection{Kinematics}
\label{sec:kinematics}

In the following subsections, we discuss the kinematic trends observed in each bubble. The presence of non-Gaussian/double-Gaussian line shapes and trends in the line velocity (Vel$_{LSR}$) and FWHM are used to examine gas kinematics.   Information about the Vel$_{LSR}$ and FWHM comes from the Gaussian line fits reported in Table \ref{LineData}.  

\begin{figure}
\begin{center}
\includegraphics[width=0.70\columnwidth]{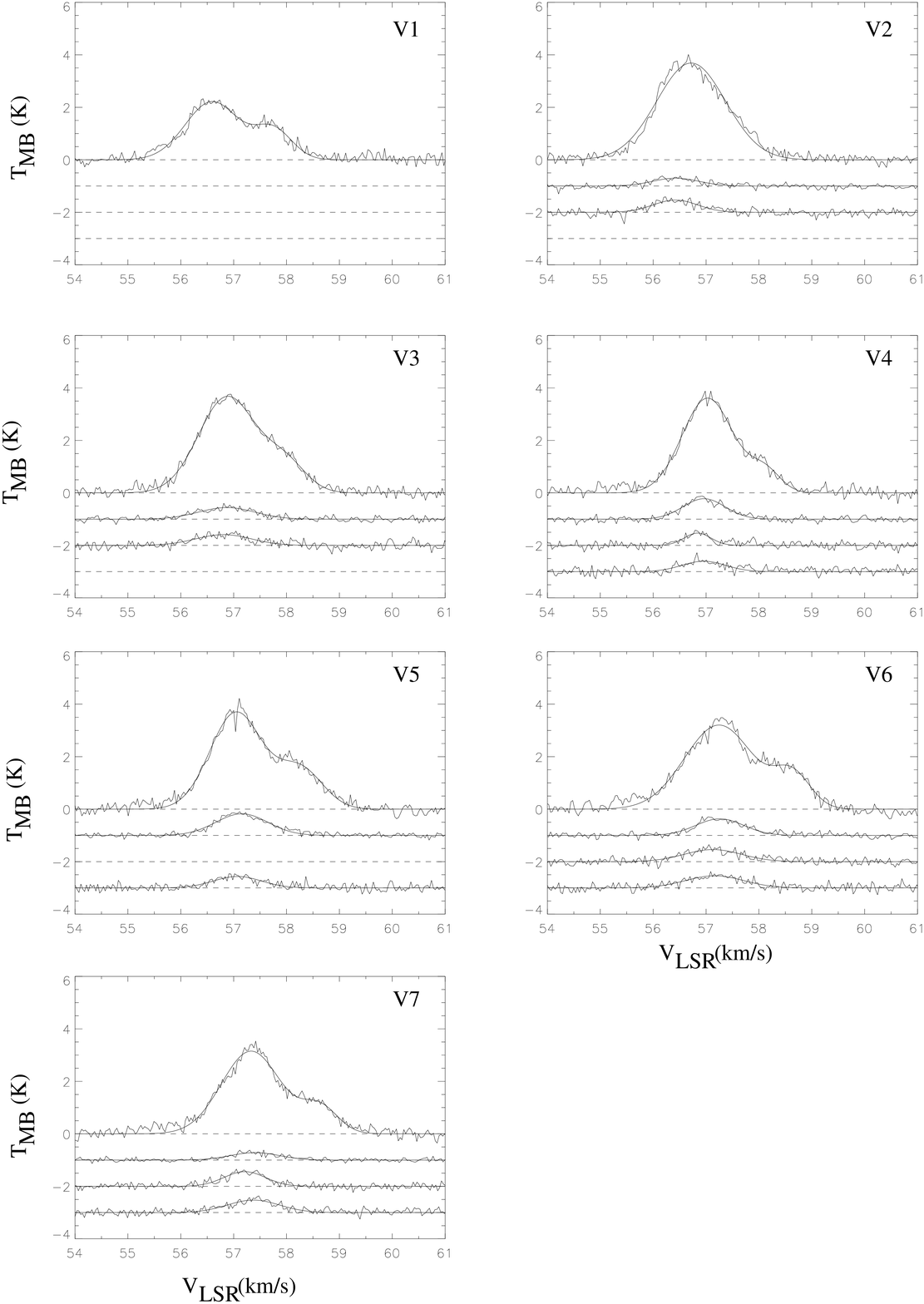}
\caption{Line profiles of detections in N62.  Only data with $>3 \sigma$ detections are plotted.  The Gaussian line fits are overplotted on the data.  The molecular lines are shown stacked and offset by 1 K.  From top to bottom, the lines shown are CS, HC$_3$N, C$^{34}$S, and CH$_3$OH.  Pointing V1 shows an example of a double-Gaussian blue-bright profile, while pointing V4 shows an example of a blue-bright ``shoulder'' asymmetry.}

\label{fig:62lineshapes}
\end{center}
\end{figure}

Whenever possible, non-Gaussian line profiles were simultaneously fit by two Gaussian profiles, indicated by consecutive lines in Table \ref{LineData}.  We also detected non-Gaussian profiles that, although best fit by a single Gaussian, showed slight asymmetries seen as ``shoulders'' in the emission line.  Some sources also had broad linewings in the CS (1-0) emission.  Representative examples of these emission line shapes are shown in Figures \ref{fig:62lineshapes} and \ref{fig:65lineshapes}. Similar double-peaked profiles, peak-and-shoulder profiles and linewings have previously been observed in studies of CS toward YSOs like \citet{Mardones97}, who examined low-mass YSOs, and \citet{olmi99}, who observed excited CS emission near H{\small II} regions. 

Double-line and ``shoulder'' profiles may be interpreted as infall if the CS (1-0) emission is optically thick; however, an infall interpretation would be supported by optically thin lines with central velocities between the double lines or peak-and-shoulder of the CS (1-0) emission \citep{Myers1996,Mardones97}.  Alternatively, optically thin lines with velocities that align with a peak in the CS (1-0) emission would support an interpretation of multiple, distinct clouds along the line of sight with similar but with slightly offset velocities.  We have assumed that all lines reported in Table \ref{LineData} except CS (1-0) are optically thin.  To test for infall, we determined the asymmetry parameter of \citet{Mardones97}, $\delta V$, which gives the difference between the velocities of the optically thick (CS) and thin (HC$_3$N, C$^{34}$S, CH$_3$OH, SiO) lines normalized by the FWHM of the optically thin line ($\Delta V_{thin}$): 

\begin{equation}
\delta V= \frac{V_{thick}-V_{thin}}{\Delta V_{thin}}
\end{equation}.

\citet{Mardones97} conducted their analysis by fitting their data with a single Gaussian fit, regardless of non-Gaussian line structure, whereas we simultaneously fit non-Gaussian lines with multiple components (Table \ref{LineData}). \citet{Mardones97} found that the CS first moment gave comparable results to their single-Gaussian fit approach. Thus, to more accurately compare our results to the \citet{Mardones97} asymmetry parameter, we used the first moment of the CS data (given in Table \ref{LineData}) as $V_{thick}$. Since the optically thin lines are all best fit by single Gaussians, we used the fit central velocities for $V_{thin}$. We report the $\delta V$ values in Table \ref{tab:deltav}. We follow \citet{Mardones97} and consider $|\delta V| > 0.25$ to be a significant indicator of gas undergoing infall (negative $\delta V$) or expansion (positive $\delta V$).

Figure \ref{fig:kinematics} shows the spatial distribution of the CS (1-0) Vel$_{LSR}$ and FWHM in each bubble. We chose CS (1-0) to map kinematic trends because it is the most common and strongest line detected in this study, and also demonstrates the spatial distribution of non-Gaussian line detections.  

We obtained additional information about the kinematics of the gas in N62, N65, and N117 using the uncalibrated data from the second, unpointed beam mentioned in Section \ref{sec:observations}. Since calibration will only influence the $T_{MB}$ values of the line, information from lineshapes, the central velocity, and FWHM of the lines is still useful for examining gas kinematics.  Since this beam was not pointed, the data from multiple pointings could not be averaged, and the signal was typically weaker.  Nonetheless, several pointings in the second beam yielded CS detections, while five pointings detected the weaker lines.  In these five pointings we conducted the $\delta V$ analysis.  The distribution of the detections from the second beam are shown in Figure \ref{fig:feed1}. 

\subsubsection{N62}

The pointings toward N62 follow an IRDC adjacent to a bubble rim. Double-Gaussian CS (1-0) line profiles were detected in every pointing except V2, although V2 showed a slight blue-bright asymmetry.  Thus non-Gaussian CS emission is seen along the full extent of the IRDC, from where it is spatially coincident with the rim to the pointings furthest from it. The blue-bright, red-dim asymmetries in all of the pointings suggest the possibility of infall, first noted for this source in \citet{Watson2016}. However, the $\delta V$ values for all of the detected optically thin lines in every pointing except V2 in N62 are below the cut-off for significant red or blue shift. In V2, the HC$_3$N and C$^{34}$S show a significant shift away from the CS peak, but blueward of the blue-bright asymmetrical CS peak, the opposite of what would be expected in an infall scenario. We note that a significant shift blueward of the CS peak is only interpreted as an expanding cloud if the CS asymmetry is red-bright, so the lineshape in V2 is not explained by expansion either.  Overall, the relative velocities of the optically thin lines compared to the CS emission supports an interpretation of two clouds along the line of sight rather than infall at these locations.  There was not a significant detection of an optically thin line at V1, closest to the bubble rim, so the reason for this double-Gaussian profile remains ambiguous.  There is a slight velocity gradient ($\sim 1$ km s$^{-1}$) from V1 to V7 and no obvious trend in linewidth variations. 

Most of the pointings with the second, uncalibrated beam detected only a single Gaussian CS profile (Fig. \ref{fig:feed1}, profile 1). There was no significant shift in velocity or FWHM of the CS lines in the second-beam map. A double-Gaussian profile appeared coincident with a bright IR feature interior to the bubble rim (Fig. \ref{fig:feed1}, profile 2).  HC$_3$N was detected in the second beam in the dark gas exterior to the bubble rim (Fig. \ref{fig:feed1}, profile 3); the $\delta V$ analysis for this pointing did not indicate infall or expansion at this location.  

\subsubsection{N65}
\label{N65}
\begin{figure}
\begin{center}
\includegraphics[width=0.70\columnwidth]{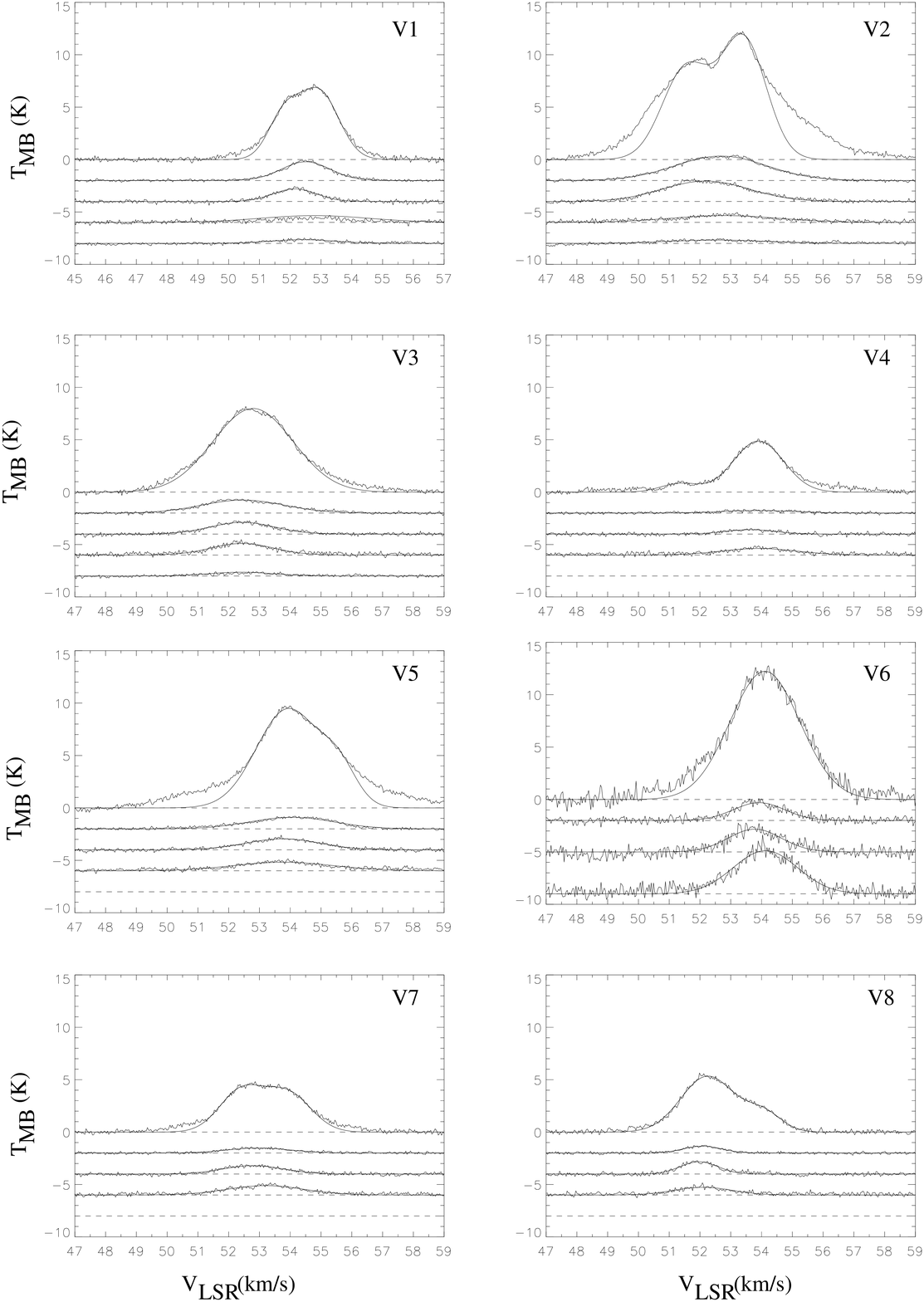}
\caption{Line profiles of detections in N65.  Only data with $>3 \sigma$ detections are plotted.  The Gaussian line fits are overplotted on the data.  The molecular lines are shown stacked and offset by 1 K.  From top to bottom, the lines shown are CS, HC$_3$N, C$^{34}$S, CH$_3$OH, and SiO.  The CS emission in pointing V2 shows a clear double-Gaussian red-bright profile with linewings, while the C$^{34}$S emission appears to have a blue-bright asymmetry. The CS and HC$_3$N emission in pointing V3 both show blue-bright asymmetries; the CS emission toward this pointing also shows linewings. In V5, the CS emission is blue-bright, but the HC$_3$N emission has a slight red-bright asymmetry.
}
\label{fig:65lineshapes}
\end{center}
\end{figure}

The pointings toward N65 trace a region previously mapped in CS (1-0) by \citet{Watson2016}.  This region is interior to the bubble rim, and contains a known YSO \citep{Watson2010} accompanied by several nearby MIR point sources.  Representative line profiles from N65 are shown in Fig. \ref{fig:65lineshapes}.  The detections of HC$_3$N, C$^{34}$S, CH$_3$OH, and CS toward N65 are notably brighter than the other sources presented here, often by an order of magnitude.  In fact, the detections of the lines assumed to be optically thin in this study (HC$_3$N, C$^{34}$S, CH$_3$OH) are bright enough that these lines may, in fact, be optically thick.  Additionally, N65 is the only source in which detectable SiO emission was observed. 

The line profiles in N65 are more complicated than those in the other sources observed in this study.  The CS emission shows both red-bright (V1, V2, V4) and blue-bright (V3,V5, V7, and V8) asymmetrical line profiles. Several of the lines detected in N65 have asymmetries despite being reasonably well fit by a single Gaussian.  The N65 V3 CS and HC$_3$N and V2 C$^{34}$S lines have blue-bright asymmetries, and the V5 HC$_3$N line is red-bright (see Fig. \ref{fig:65lineshapes}).  Interestingly, the CS emission in pointing V2 is red-bright, while the C$^{34}$S emission appears blue-bright, and in pointing V5 the CS emission is blue-bright, but the HC$_3$N emission has a slight red-bright asymmetry; both cases may be a result of the strong linewings. Linewings were detected in the CS lines toward all of the pointings in N65 except V8.  A significant $\delta V$ offset was only detected toward V8. As seen in N62, the HC$_3$N and C$^{34}$S show a significant shift away from the CS peak, but blueward of the blue-bright asymmetrical CS peak, inconsistent with an infall model.

\begin{figure}
\begin{center}
\includegraphics[width=0.70\columnwidth]{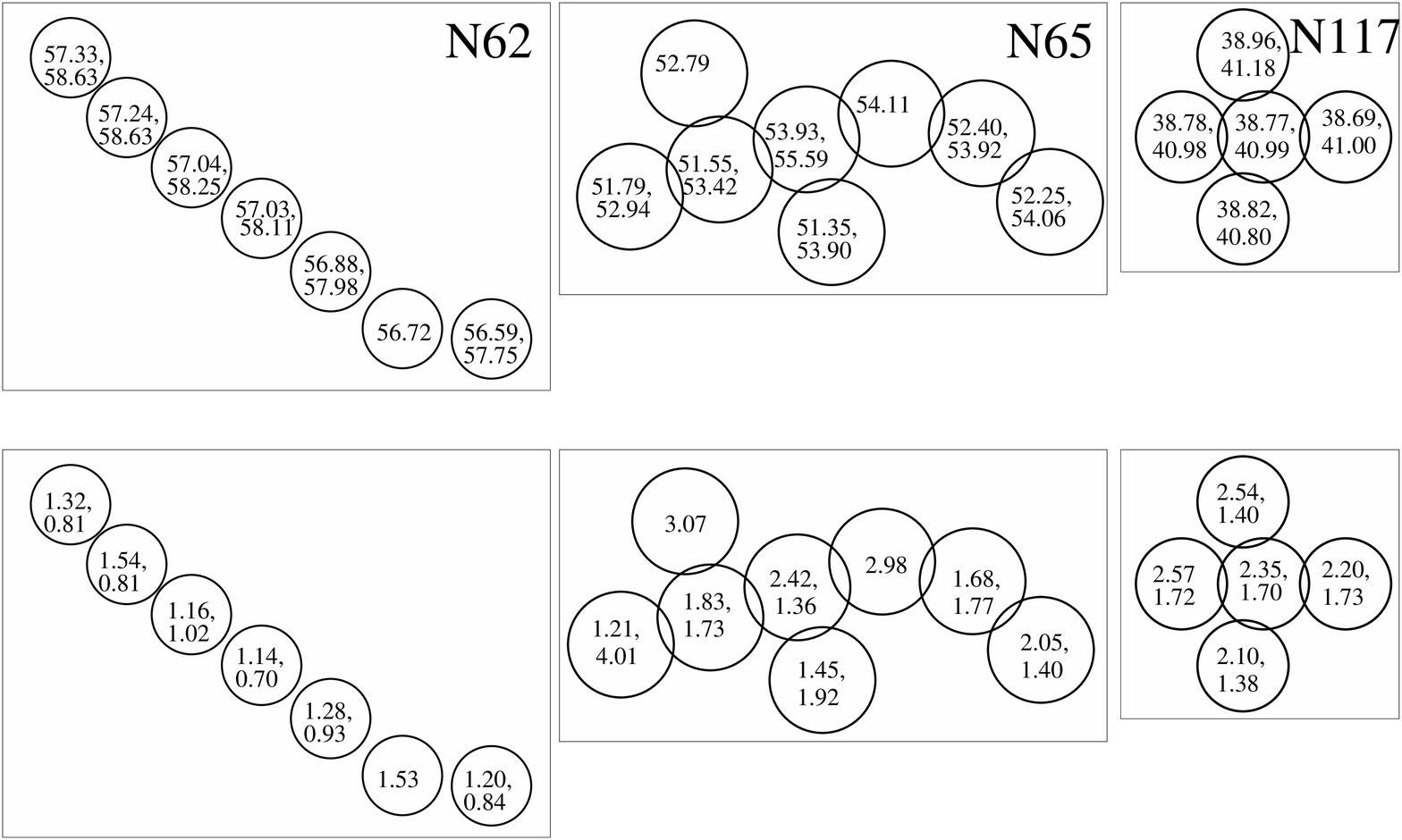}
\caption{Distribution of the CS (1-0) Vel$_{LSR}$ (top row) and FWHM (bottom row) values reported in Table \ref{LineData} for N62, N65, and N117.  Circles represent the pointings shown in Figure \ref{fig:bubbles}.  All values have units of km s$^{-1}$.  Two values indicate that a double-Gaussian profile was fit at that location.
}
\label{fig:kinematics}
\end{center}
\end{figure}

\citet{Watson2016} noted red-bright asymmetric line profiles and a velocity gradient of $\sim 1$ km s$^{-1}$ in the area covered by this study.  The gradient in CS (1-0) velocities was confirmed in this study and is shown in Fig. \ref{fig:kinematics}; the velocity gradient was also detected in the velocities of the optically thin lines.  

The data from the second, uncalibrated beam further support a multiple-cloud interpretation of the data and show that N65 has the most interesting chemistry of the bubbles examined in this study. A majority of the CS detections show double-Gaussian and asymmetric profiles (Fig. \ref{fig:feed1}, profiles 5-8). All four pointings with optically thin line detections align with the IR dark feature interior the bubble rim, while the pointings along the rim detect only CS. The $\delta V$ analysis on the four detections of optically thin lines only indicated $\delta V > 0.25$ in the location shown in Figure \ref{fig:feed1}, profile 7, which had a $\delta V = 2.6$.  However, as Fig. \ref{fig:feed1}, profile 7 shows, the HC$_3$N closely aligns with the weaker CS line rather than falling between the two peaks as would be expected in a profile caused by infall.  Using $\delta V > 0.25$ to indicate infall in the \citet{Mardones97} model assumes the optically thin lines are associated with the brighter component of a double-Gaussian peak in the optically thick emission.  The HC$_3$N association with the weaker CS peak in this pointing does not fit the \citet{Mardones97} infall model and is instead further evidence of interesting chemistry in N65.    



\subsubsection{N117}

\begin{figure}
\begin{center}
\includegraphics[width=0.70\columnwidth]{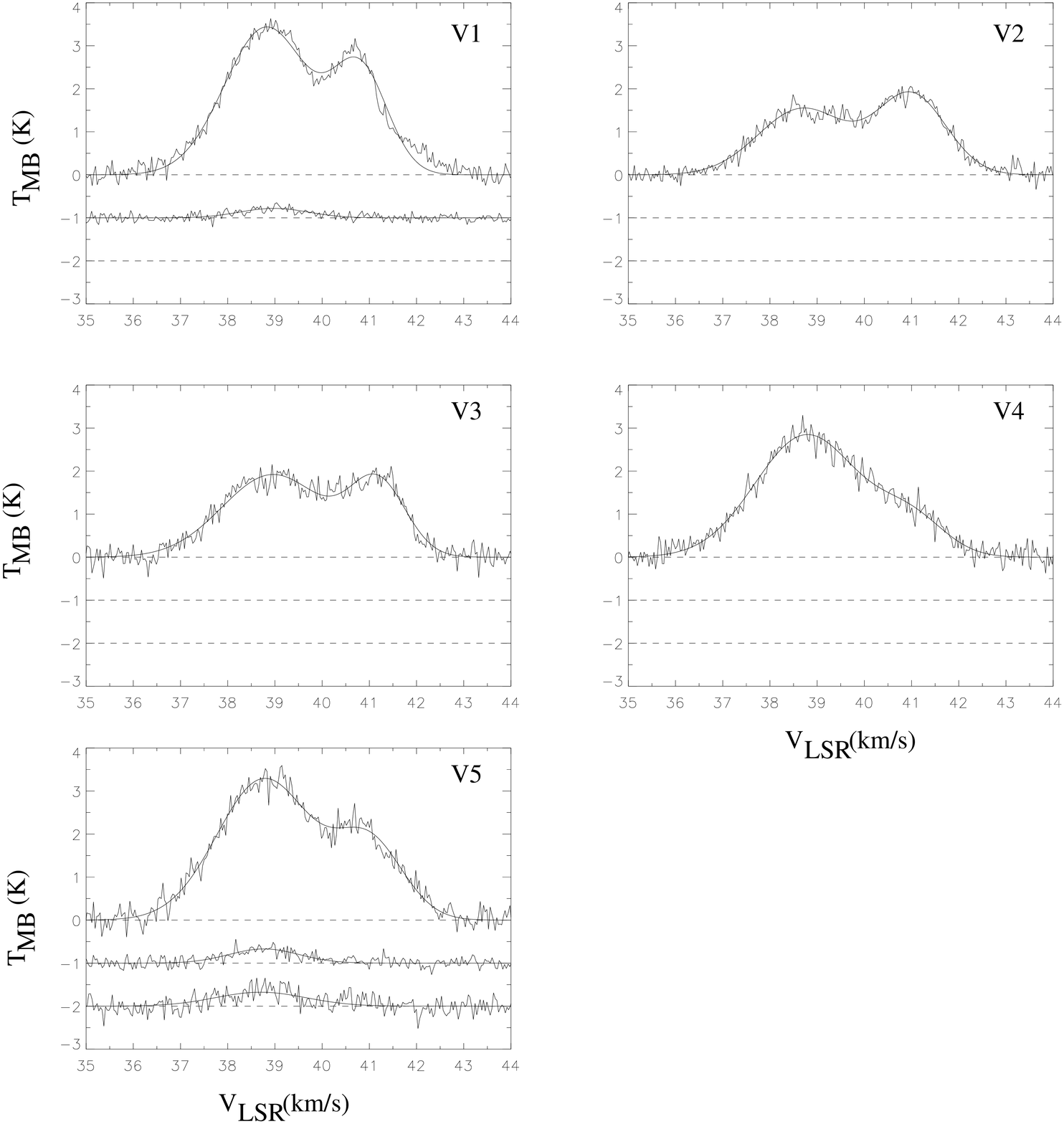}
\caption{Line profiles of detections in N117.  Only data with $>3 \sigma$ detections are plotted.  The Gaussian line fits are overplotted on the data.  The molecular lines are shown stacked and offset by 1 K.  From top to bottom, the lines shown are CS, HC$_3$N, and C$^{34}$S. The emission in pointing V1 shows a blue-bright double-Gaussian, V2 shows a red-bright double-Gaussian, and V3 shows a double-Gaussian with peaks of comparable brightness.
}
\label{117lineshapes}
\end{center}
\end{figure}

The pointings in N117 are toward and next to a YSO inside the bubble rim. In \citet{Watson2016}, observations toward the YSO indicated a blue-bright line asymmetry possibly indicative of infall.  Our results detect double-Gaussian profiles in all five pointings toward N117; however, in pointing N117 V2, the red component is brighter than the blue, and in N117 V3 the red and blue components have comparable brightness, as shown in Fig. \ref{117lineshapes}.  In V5, the $\delta V$ values of HC$_3$N and C$^{34}$S show a significant shift away from the CS peak, but like N62 and N65  this shift is blueward of the blue-bright asymmetrical CS peak. Combined, the $\delta V$ values and presence of both blue-bright and red-bright asymmetric profiles indicates multiple cloud structure rather than infall.

\begin{figure}
\begin{center}
\includegraphics[width=0.70\columnwidth]{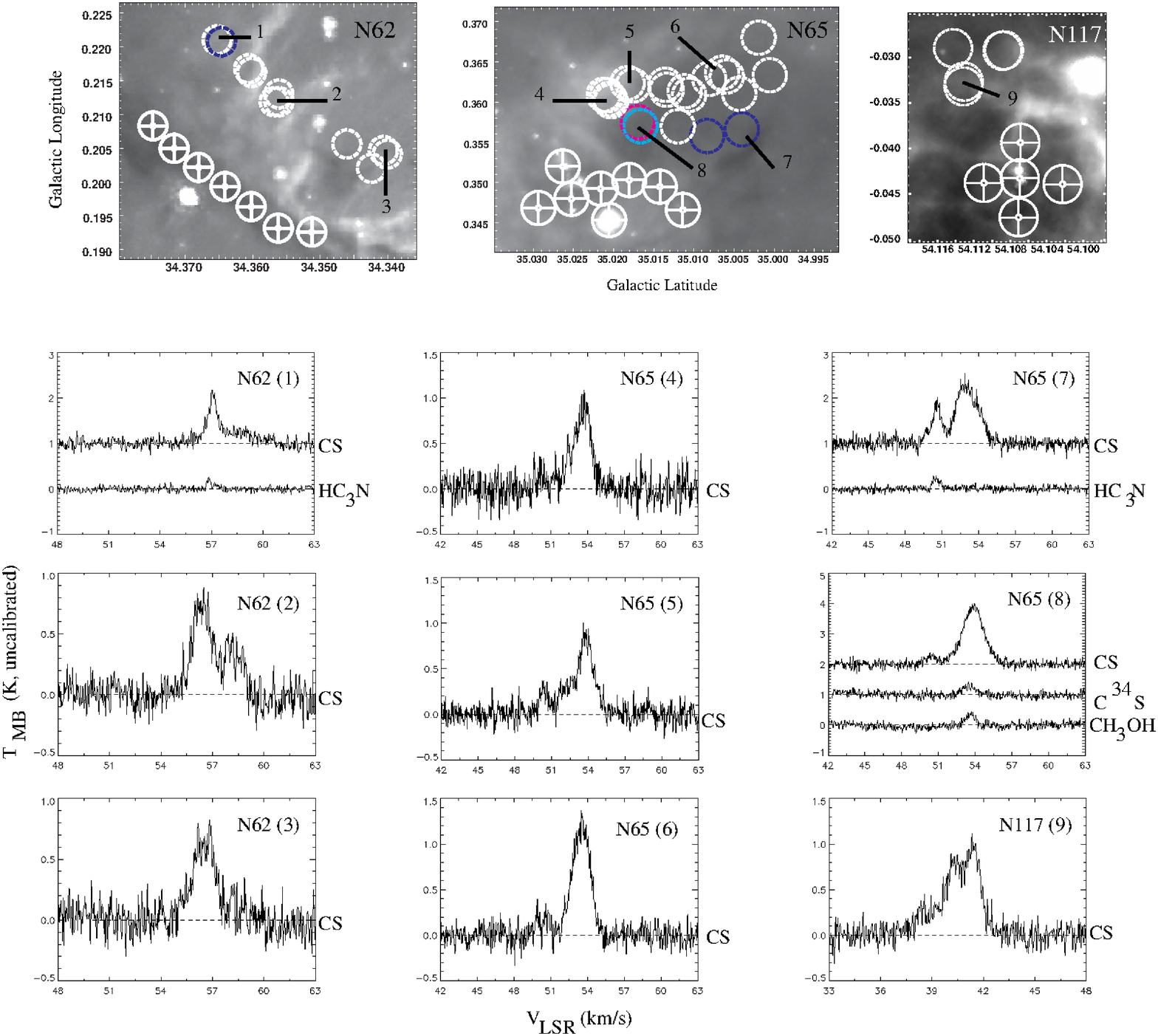}
\caption{Distribution and representative profiles of detections in the second (uncalibrated) beam. The grey scale images show 8 $\mu$m emission, zoomed in on regions shown in Figure \ref{fig:bubbles}. Circles with interior crosshairs indicate the pointings shown in Table \ref{tab:pointings} and Figure \ref{fig:bubbles}. Dashed circles indicate $> 3\sigma $ detections in the second beam. For the dashed circles, line color corresponds to the following detections: white (CS only), dark blue (CS and HC$_3$N), cyan (CS and C$^{34}$S), and magenta (CS, C$^{34}$S, and CH$_3$OH).The numbered locations on the 8 $\mu$m images correspond to the representative profiles shown below. In pointings where multiple lines were detected, the profiles are shown stacked and offset by increments of $1$ K for clarity.
}
\label{fig:feed1}
\end{center}
\end{figure}

A representative profile detected in the second, uncalibrated beam is shown in Fig. \ref{fig:feed1}. The second beam detected a red-bright asymmetric line, similar to that seen pointing V2.  This result supports the interpretation of two clouds at offset velocities that extend past the area sampled by our pointings. 

\begin{table}
\caption{Normalized asymmetry parameter $\delta V$ for observed optically thin lines.}
\label{tab:deltav}
\begin{tabular}{lcccc}
\hline
Source & $\delta V_{HC_3N}$ & $\delta V_{C^{34}S}$ & $\delta V_{CH_3OH}$ & $\delta V_{SiO}$ \\
\hline
N62 V2 & \bf{0.35}& \bf{0.36} & $\cdots$ & $\cdots$\\
N62 V3 & 0.15& 0.20 & $\cdots$ & $\cdots$\\
N62 V4 & 0.00  & 0.20 & 0.00 & $\cdots$ \\
N62 V5 & 0.08& $\cdots$ &0.17 & $\cdots$\\
N62 V6 & -0.02 & 0.08 &0.06 & $\cdots$\\
N62 V7 & 0.00 & 0.20 & 0.02 & $\cdots$\\
\hline
N65 V1 & -0.02 & 0.20  &-0.07 & 0.02 \\
N65 V2 & 0.06 & 0.21 &0.023 &  0.09 \\
N65 V3 & 0.13 & 0.18 &0.19 & 0.20 \\
N65 V4 & -0.07 &0.03 &-0.08 & $\cdots$\\
N65 V5 & 0.05 & 0.16 & 0.12 & $\cdots$ \\
N65 V6 & 0.07 &0.14 &-0.02 & $\cdots$\\
N65 V7 & 0.07 & 0.15 &-0.02 & $\cdots$\\
N65 V8 & \bf{0.36} & \bf{0.44} & 0.23 & $\cdots$\\
\hline
N117 V1 & 0.24 & $\cdots$ & $\cdots$ & $\cdots$\\
N117 V5 & \bf{0.32} & \bf{0.31} & $\cdots$& $\cdots$ \\
\end{tabular}

\medskip
\textbf{Notes.} Normalized asymmetry parameter $\delta V$ for each observed optically thin line. Negative values indicate the optically thin line was red-shifted relative to the brightest peak in the optically thick line.  Lines with a significant shift ($|\delta V| > 0.25$) are emphasized in bold.
\end{table}

\section{Discussion: How are bubbles influencing their surroundings?}
\label{sec:discussion}

We had hypothesized that bubble rims coincident with IRDCs would show signs of infall in the IRDC gas nearest the rim, and lack of infall away from the rim.  Our findings do not support this hypothesis.  Although the molecular lines that we detect are associated with dense, hot, and/or shocked gas possibly linked to star formation, the kinematics of the gas do not indicate infall.

Based on comparisons between the velocities of our dense gas detections and radio recombination line detections toward the bubbles \citep{anderson2011}, our detections toward N90 are likely not associated with the bubble.  On the other hand, bubble N62 is likely associated with nearby dense gas, but we failed to detect compelling evidence of infall in N62, and instead found that in pointings with non-Gaussian CS emission, optically thin line detections typically more closely align in velocity with the brightest peak of CS. This finding supports non-Gaussian CS lines being the product of multiple clouds, close but slightly offset in velocity, along the line of sight.  We note that while there is evidence of star formation in  IRDC G34.43+0.24 \citep{xu2016, rathborne2005}, our findings focus on collapse of the IRDC in the area immediately adjacent to N62.  Multiple cloud components at close velocities, or substructure, has been previously detected in IRDCs and may be a common feature of such objects (e.g. \citet{Devine2011, Dirienzo2015}). This multiple-clouds interpretation may also explain the double peaked, asymmetric profiles seen elsewhere in IRDC 34.43+0.24 by \citet{xu2016}. Compared to the gas interior to bubble rims in N65 and N117, the linewidths of the gas in the IRDCs associated with N62 are narrower and there is less variation in the central velocities of the lines. This kinematic structure indicates that the IRDC gas is more quiescent than the gas interior to bubbles, further supporting lack of IRDC disruption by the bubbles.  

N65 and N117 were observed to examine the gas within the bubble.  In both N65 and N117, we have assumed that the gas is interior to the bubble rims, although it is possible that the gas is exterior to the bubble rim.  In this case, evidence of disrupted gas may be a sign of the bubble impacting exterior gas rather than the effect of the bubble rim passing by interior gas.  However, we do not detect signs of increased disruption in the gas along the bubble rim in N62, which would be expected if we were observing a bubble rim's impact on exterior gas.  We therefore believe that the gas observed in N65 and N117 is more likely interior to the bubble rims.  Wider linewidths and larger velocity gradients than seen in the IRDCs indicates that the gas is relatively more disrupted. The nature of the bubbles' impact on the gas, however, is difficult to discern due to the complicated kinematics detected in the gas within the bubbles.  

In \citet{Watson2016} the blue-bright non-Gaussian line profile of N117 was interpreted as infall, but observations of surrounding gas in the clump as well as optically thin lines suggest that the infall interpretation of the profile is not correct. As in the N62 IRDC observations, the velocities of the optically thin lines align with the peak of the brightest component of the CS line.  All five pointings in N117 show double peaks which have a fairly consistent velocity separation but vary in brightness, such that at some of the pointings the double peaks are red-bright, while others are blue-bright. It is unlikely that infall and expansion sources at different pointings near the cloud would maintain such consistent velocity separations, suggesting multiple clouds of gas moving at slightly different velocities as we have proposed for the IRDC adjacent to N62. The  massive stellar outflow candidates identified by \citet{cyganowski2008} at pointings V1 and V5 may be playing a role in generating the kinematic structure of N117, although the nature of this role is beyond the scope of this paper.

A multiple-cloud interpretation was suggested in \citet{Watson2016} as a possible explanation for the red-bright non-Gaussian structure detected in N65. Our observations lend support to this hypothesis, but also have found the presence of CS linewings in all but one of the pointings toward N65, suggesting a possible cloud-scale outflow. Further evidence for outflow in N65 is the detection of thermal SiO emission, an outflow tracer \citep{Bachiller96}.  Our data indicate that the kinematics of the gas interior to N65 is disrupted, but a more detailed analysis of the kinematics requires higher resolution observations and a full map of the gas to better measure velocity gradients and outflow extents.  The complicated kinematics of this region are consistent with it being near a trapezium-like protocluster as suggested by \citet{brogan2011}. We also note that the line strengths of all the detected lines are much stronger in N65 than in the other sources, indicating a possible interesting chemical formation pathway to explore in further work. The high line strengths of the molecular gas in N65 make this source the best candidate for such follow-up observations.

\section{Conclusions}
\label{sec:conclusions}

We used the GBT to observe $21$ emission lines between $39$ GHz and $49$ GHz toward $26$ pointings in four MIR bubbles to explore the interaction between bubbles and the surrounding ISM. Our sources were selected to explore two MIR bubbles spatially coincident with IRDCs, and two bubbles with evidence of YSOs interior to the bubble rims. All four sources had shown asymmetric CS (1-0) line profiles in previous work \citep{Watson2016}. Our conclusions are as follows:

\begin{itemize}

\item We successfully detected $> 3 \sigma$ emission from five molecules: CS, C$^{34}$S, CH$_3$OH, HC$_3$N, and SiO. CS (1-0) was detected in all but two pointings.  C$^{34}$S, CH$_3$OH, HC$_3$N were detected in N62, N65, and N117. SiO was detected only in N65.  

\item Based on the velocity differences of the dense gas tracers detected in this study and \citet{Watson2016}  compared to the  radio recombination lines detected in N90 by \citet{anderson2011}, the dense gas we detected is not associated with bubble N90.

\item We calculated the column densities for each of the detected lines, and used publicly available Hi-GAL data to determine the total column density and abundances of the detected molecules. We examined the relative isotopic abundance of [C$^{34}$S]/[CS] and found a value of $\sim 0.1$, higher than the typically assumed value of $0.045$ but consistent with our abundance ratio being an upper limit due to self-absorption in the CS line. CS abundances were typically higher interior to bubble rims than in the IRDCs by a factor of a few. The detections of the other lines were not consistent enough to determine any trends; however, the linestrengths and abundances in N65 were significantly higher than in the other sources. 

\item  We used Gaussian line fits to the data to examine linewidths and central velocities of the lines. Several CS detections showed non-Gaussian line profiles as either double-peaked emission or asymmetric (i.e. ``shoulder'') lines. These lines were simultaneously fit with two Gaussian lines to best fit the data. Due to the presence of complicated line structure and linewings, the first moment of the CS emission was also determined.  The line fits were used to examine the velocity gradients and linewidths in the sources. The lines were narrower and showed less velocity variation in the IRDCs relative to the pointings within the bubbles.

\item We compared the first moment of CS profiles relative to the velocities of the other detected lines using the \citet{Mardones97} $\delta V$ parameter to look for evidence of infall. We did not detect evidence of infall in any of the pointings.

\item Our results toward bubble N62 suggest that the bubble rim's coincidence with an IRDC has not caused collapse in the IRDC measurable by infall line profiles.  In N65 and N117, the velocity structure of gas interior to bubble rims shows evidence of disruption by the bubble.

\end{itemize}   

Our results leave a key question unanswered: When does an interaction with a bubble change an IRDC's dynamics? Future studies with a wider sample of bubble rim - IRDC interactions may yield a better understanding of collapse and infall. Furthermore, our results indicate that understanding of the dynamics of the gas within bubbles requires higher resolution maps of the gas. We recommend N65 as a target for such studies based on its interesting kinematics and relatively strong molecular line emission.

The authors thank the anonymous referee for useful comments and feedback that improved this document.

\software{GBTIDL \citep{GBTIDLref}, 
 SAOImage DS9 \citep{ds9ref}}

\bibliographystyle{aasjournal}


\end{document}